\documentclass[prl,aps,twocolumn,superscriptaddress,longbibliography]{revtex4-2}

\usepackage{amsmath}
\usepackage{amssymb}
\usepackage{calligra}
\usepackage{graphicx}
\usepackage{hyperref}
\usepackage{color}

\makeatletter

\newif\ifinsupplement
\let\revtex@addcontentsline\addcontentsline

\renewcommand{\addcontentsline}[3]{%
  \ifinsupplement
    \def\revtex@ext{#1}%
    \def\revtex@toc{toc}%
    \ifx\revtex@ext\revtex@toc
      \revtex@addcontentsline{stoc}{#2}{#3}%
    \else
      \revtex@addcontentsline{#1}{#2}{#3}%
    \fi
  \else
    \revtex@addcontentsline{#1}{#2}{#3}%
  \fi
}

\newcommand{\supplementtableofcontents}{%
  \begingroup
    \let\revtex@starttoc\@starttoc
    \def\@starttoc##1{%
      \def\revtex@ext{##1}%
      \def\revtex@toc{toc}%
      \ifx\revtex@ext\revtex@toc
        \revtex@starttoc{stoc}%
      \else
        \revtex@starttoc{##1}%
      \fi
    }%
    \def\addcontentsline##1##2##3{}%
    \tableofcontents
  \endgroup
}

\newcommand{\StartSupplementTOCEntries}{\global\insupplementtrue}
\newcommand{\StopSupplementTOCEntries}{\global\insupplementfalse}

\makeatother
\newcommand{\bdy}[0]{\mathsf{bdy}}
\newcommand{\bulk}[0]{\mathsf{bulk}}

\newcommand{\covariance}[0]{\mathcal{V}}
\newcommand{\symplectic}[0]{\mathcal{S}} % because we also have entropy, it is important to keep this S different
\newcommand{\ximat}[0]{\ensuremath{\boldsymbol{\xi}}}
   
\newcommand{\avg}[1]{\ensuremath{\langle #1 \rangle}}

\newcommand{\Bdy}[0]{\covariance_\bdy}
\newcommand{\Bulk}[0]{\covariance_\bulk}
\newcommand{\Corr}[0]{\mathcal{C}}

\newcommand{\diff}[1]{{#1}} % Use this command to render without highlighting diffs

\newcommand{\SEE}{\ensuremath{S}}
\newcommand{\Tr}{\ensuremath{\mathrm{Tr}}}
\newcommand{\Sym}{\ensuremath{\mathcal{S}}}
\newcommand{\Sform}{\ensuremath{\Omega}}
\newcommand{\Cmat}{\ensuremath{\mathcal{V}}}
\newcommand{\Seig}{\ensuremath{\mathcal{\nu}}}
\newcommand{\depth}{\ensuremath{R}}

\newcommand{\lads}{\ell_{\text{AdS}}}
\newcommand{\db}{d_{\text{bulk}}}
\newcommand{\cov}{\ensuremath{\mathrm{Cov}}}
\newcommand{\var}{\ensuremath{\mathrm{Var}}}
\newcommand{\corr}{{\mathrm{Corr}}}

\DeclareMathAlphabet{\mathcalligra}{T1}{calligra}{m}{n}
\DeclareFontShape{T1}{calligra}{m}{n}{<->s*[2.2]callig15}{}
\newcommand{\scriptr}{\mathcalligra{r}\,}

\makeatletter
\newcommand{\MainRef}[1]{%
  \@ifundefined{r@MAIN-#1}{\ref{#1}}{\ref{MAIN-#1}}%
}
\newcommand{\MainEqref}[1]{%
  \@ifundefined{r@MAIN-#1}{\eqref{#1}}{\eqref{MAIN-#1}}%
}
\makeatother

\newcommand{\mcite}[1]{\cite{#1}}

\newcommand{\SMCombinedTitle}[1]{%
  \par\clearpage
  \begingroup
    \centering
    \normalfont
    \vspace*{6pt}%
    {\large\bfseries #1\par}%
    \vspace*{11.5pt}%
  \endgroup
}

\begin{document}

% --- Main body: define both \label{X} and \label{MAIN-X} ---
\begingroup
\let\oldlabel\label
\renewcommand{\label}[1]{%
  \oldlabel{#1}%
  \oldlabel{MAIN-#1}%
}
\title{Building Holographic Entanglement by Measurement}

\author{Jonathan Jeffrey}
\affiliation{Department of Physics, Stanford University, Stanford, California 94305, USA}
\affiliation{Department of Applied Physics, Stanford University, Stanford, California 94305, USA}

\author{Lucien Gandarias}
\affiliation{Department of Physics, Stanford University, Stanford, California 94305, USA}

\author{Monika Schleier-Smith}
\affiliation{Department of Physics, Stanford University, Stanford, California 94305, USA}

\author{Brian Swingle}
\affiliation{Department of Physics, Brandeis University, Waltham, Massachusetts 02453, USA}

\date{\today}

\begin{abstract}
We propose a framework for preparing quantum states with a holographic entanglement structure, in the sense that the entanglement entropies are governed by minimal surfaces in a chosen bulk geometry. We refer to such entropies as holographic because they obey a relation between entropies and bulk minimal surfaces, known as the Ryu-Takayanagi (RT) formula, that is a key feature of holographic models of quantum gravity. Typically in such models, the bulk geometry is determined by solving Einstein's equations. Here, we simply choose a bulk geometry, then discretize the geometry into a coupling graph comprising bulk and boundary nodes. Evolving under this graph of interactions and measuring the bulk nodes leaves behind the desired pure state on the boundary. We numerically demonstrate that the resulting entanglement properties approximately reproduce the predictions of the Ryu-Takayanagi formula in the chosen bulk geometry. We consider graphs associated with hyperbolic disk and wormhole geometries, but the approach is general. \diff{We also introduce a procedure for decorating the graph which, when applied to the hyperbolic disk, yields a genuine CFT ground state with power-law correlations but which only partially obeys RT.} The minimal ingredients in our proposal involve only Gaussian operations and measurements and are readily implementable in photonic and cold-atom platforms.
\end{abstract}

\maketitle

At the largest length scales, gravity dominates, but at the smallest scales, the rules of quantum theory hold sway. Physicists have long searched for a more general framework that combines these two regimes, a theory of quantum gravity. Direct experimental access to quantum effects in gravity has not yet been achieved, so the study of quantum gravity is driven by a number of conceptual puzzles that arise when quantum field theory and general relativity are combined. For example, the realization that black holes are thermal objects with an entropy proportional to their area rather than their volume~\cite{Bekenstein1973,Hawking1974} led to the idea that quantum gravity is fundamentally holographic, with the geometry of spacetime emerging from microscopic degrees of freedom living on a boundary of one dimension fewer~\cite{Susskind:1994vu,tHooft:1993dmi}. The best understood version of this holographic framework is known as the anti-de Sitter space / conformal field theory correspondence (AdS/CFT)~\cite{Maldacena:1997re,Witten:1998qj,Gubser:1998bc,Aharony:1999ti}. This correspondence is a precise duality between a quantum system with no dynamical gravity (CFT) and a quantum gravitational system (AdS). The CFT is naturally defined at the boundary of the dual ``bulk'' asymptotically AdS spacetime. 

Entanglement plays a central role in AdS/CFT and is closely related to the emergence of spacetime~\cite{ryu_holographic_2006,VanRaamsdonk2010}. The AdS/CFT dictionary relates states in the boundary to states in the bulk, which often correspond to smooth geometries. In such cases, the bulk geometry encodes the entanglement structure of the CFT as quantified by the Ryu-Takayanagi (RT) formula: the von Neumann entropies of subregions of the boundary CFT are equal to the areas of corresponding minimal surfaces in the bulk geometry, to leading order in Newton's constant \cite{ryu_holographic_2006}. This formula can be thought of as a vast generalization of the older formula for black hole entropy~\cite{Bekenstein1973,Hawking1974}, to which it reduces in special cases. 

Holographic duality is less well understood outside of the small set of special quantum systems whose bulk duals correspond to some form of Einstein gravity in the semiclassical limit~\cite{Aharony:1999ti}. A core issue is that the boundary side is in general a strongly correlated system that is hard to analyze with conventional theoretical tools. Yet the boundary is potentially amenable to quantum simulation, as suggested by several theoretical proposals~\cite{swingle2016measuring, pikulin2017black, bentsen2019treelike, yoshida2019disentangling, brown2023quantum, dey2024simulating, cowsik2025engineering, sahay2025emergent} and proof-of-principle experiments~\cite{li2019measuring, landsman2019verified, blok2021quantum, periwal2021programmable, Jafferis:2022crx,Shapoval2023towardsquantum, granet2025simulating} aimed at simulating aspects of holographic duality in the lab.  \diff{Whereas standard holographic theories have many co-occurring phenomena, such as holographic entanglement and strong chaos, quantum simulations offer a route to exploring which of these aspects can exist independently or controllably combining them.  For investigating entanglement, a crucial challenge is preparing holographic boundary states for arbitrary bulk geometries.}

In this paper, we show how states with holographic patterns of entanglement can be prepared using programmable interactions and measurements in constant time. We focus on a Gaussian version of our protocol for its tractability, but we also discuss prospective generalizations. The key idea is illustrated in Figure~\ref{fig:setup}: given a desired geometry, which is discretized into a graph, one first evolves an initially unentangled state with a simple quench Hamiltonian determined by the graph.  One then measures the bulk part of the graph, leaving behind a pure state on the boundary. Our central result is that this boundary state has a holographic entanglement structure in that the entropies of boundary subregions are approximately equal to the predictions of the RT formula in the chosen geometry. Our framework provides both a theoretical laboratory for further investigation of holographic entanglement and an experimentally accessible method for engineering designer bulk geometries. In particular, photonic systems, atomic ensembles in cavities, or superconducting circuits can be used to implement our protocol with existing technology.

\begin{figure*}
    \centering
    \includegraphics[width=\linewidth]{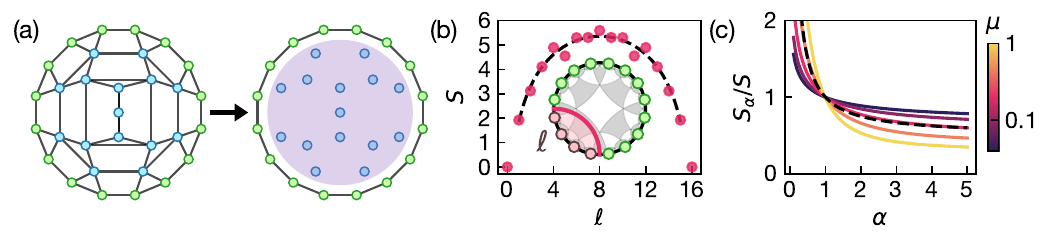}
    \caption{(a) The quench-and-measure protocol. The input state is a set of independent squeezed Gaussian modes, with some designated as the bulk (blue circles) and others as the boundary (green circles). After quenching on interactions with a specified coupling graph (black lines), chosen here to discretize a hyperbolic disk, measurement of the bulk nodes prepares the boundary in a pure state. (b) Entanglement entropies $S$ of connected boundary regions of length $\ell$ for initial squeezing
    parameter $\mu=0.2$, compared with the fit to the prediction for a (1+1)D CFT on a circle (dashed black line), where the central charge $c=6.5(3)$ and vertical offset $\epsilon = 5.36(8)$ are free parameters. Through the Ryu-Takayanagi formula, the entanglement entropy $S$ has a geometrical interpretation as proportional to the minimal length in a holographic bulk (inset schematic). (c) R{\'e}nyi entropy (normalized by entanglement entropy) vs index $\alpha$ for a region size $\ell = 4$ on the boundary and various squeezing parameters $\mu$ (colored lines), compared to the CFT prediction (dashed black line).
    }
    \label{fig:setup}
\end{figure*}

Our approach is partly inspired by tensor-network models of holographic duality~\cite{swingle2012entanglement,happy_2015,rtn_2016} and measurement-based protocols for preparing entangled states~\cite{briegel2001persistent,tantivasadakarn2024long,cowsik2025engineering}. We have focused on the RT formula as the primary characteristic of holographic entanglement, but a full holographic dual has additional properties. These include R{\'e}nyi entropies which depend non-trivially on the R{\'e}nyi index and power-law correlations for some observables. Many existing toy models of holographic entanglement are unable to capture these additional features, for example, tensor network models based on stabilizer states~\cite{happy_2015}, as well as random tensor networks~\cite{rtn_2016}, have R{\'e}nyi entropies that are independent of the R{\'e}nyi index and no power-law correlations. Since these early examples, a wide variety of tensor network and holographic code models have been developed, some of which also have power-law correlations~\cite{latorre2015holographiccodes,Bhattacharyya_2016,Donnelly_2017,Han_2017,Harris_2018,Jahn_2019,hung2019padiccftholographictensor,Osborne_2020,gesteau2020infinitedimensionalhappycodeentanglement,Cao_2021,Jahn_2022,Cao_2022,bao2025qgsymqrgads3cft2correspondence,Jahn_2021}. 

\diff{Given a target bulk geometry, we consider two different kinds of graphs, ``undecorated'' and ``decorated.'' In the undecorated case, we obtain entropies which can be made arbitrarily large, a non-trivial Renyi spectrum, but no clean power-law correlations. Consistent with this, we find that these states are not genuine CFT states despite approximately reproducing the RT formula in a variety of geometries. By contrast, in the decorated case for a hyperbolic geometry, we find power-law correlations but agreement with RT only for single intervals. We offer considerable evidence that the decorated case yields the ground state of a regulated $c=1$ free scalar CFT.}

\textit{Quench-and-Measure Protocol}---We consider a collection of $N$ sites, each of which is host to a single position degree of freedom $x_i$ with an associated canonical momentum $p_i$. These quadrature operators obey the standard commutation relations, $[x_i , p_j] = i \delta_{ij}$, for $i,j \in \{1,\cdots,N\}$. It will be convenient to group these quadrature operators into a phase space vector, $\ximat = (x_1, \ldots, x_N, p_1,\ldots,p_N)$. For later use, we also divide these degrees of freedom into two sets: \diff{the boundary ($\bdy$), on which we aim to prepare a holographic state; and the bulk ($\bulk$), comprising ancillae that will facilitate the state preparation and subsequently be discarded.}

The system is initialized into an unentangled Gaussian state of the form
\begin{equation}
    \psi_0(x) \propto \prod_i e^{- \frac{\mu}{2} x_i^2}.
\end{equation}
Here, $\mu = 1$ represents the case where each oscillator is initialized in the vacuum state, and more generally $\mu \neq 1$ allows for initial squeezing in each mode. Starting from this initial state, we quench the system by evolving for a time $t$ with the Hamiltonian 
\begin{equation}
    H_q = \frac{1}{2}\sum_{i,j} J_{ij} x_i x_{j},
\end{equation}
where $J$ is a coupling matrix between the oscillators. $J$ must be real and symmetric, and in the \diff{undecorated} case it will be the adjacency matrix of an unweighted graph.  For strong initial squeezing $\mu \ll 1$ (or long time $t$), the resulting entangled state is then a continuous-variable graph state~\cite{menicucci2006universal}, and its entropies are closely related to those of ``perfect tensors’’~\cite{kwon_most_2025}.  More generally, the entanglement depends only on the combination $t/\mu$, so below we will fix the quench time to $t=1$.

Following the quench, we measure all the bulk sites in the momentum basis. Based on the measurement outcomes, we may optionally apply local phase space displacements to the unmeasured boundary sites to obtain a state in which $\langle x_i \rangle = \langle p_i \rangle = 0$ for all $i \in \bdy$. The resulting state is a pure state of the boundary degrees of freedom that does not depend on the measurement outcome. \diff{The bulk degrees of freedom are removed by the measurement and no longer participate in the state.} We will investigate the entanglement structure of the resulting boundary state for different choices of the coupling matrix $J$ and initial squeezing $\mu$.

The entanglement between the modes of the Gaussian state can be determined from the covariance matrix $\covariance = \avg{\ximat \ximat^T} - \avg{\ximat}\avg{\ximat}^T$, with elements $\covariance_{ij} = \mathrm{Cov}(\xi_i, \xi_j) = \langle \left\{ \xi_i - \langle{\xi_i}\rangle, \xi_j - \langle{\xi_j}\rangle \right\}\rangle $~\cite{holevo1999capacity,eisert2010colloquium,weedbrook2012gaussian,supp}.  We thus focus on how the covariance matrix transforms under the quench and under measurements. For initial squeezing $\mu$, the initial covariance matrix is
\begin{equation}
\covariance_0 =
\begin{pmatrix}
\covariance_{0, x} & 0 \\
0 & \covariance_{0, p}
\end{pmatrix}
=
\begin{pmatrix}
\frac{1}{2\mu} I & 0 \\
0 & \frac{\mu}{2} I
\end{pmatrix}
\end{equation}
where $I$ is the identity matrix.  Solving the Heisenberg equation of motion $d\ximat/dt = i[H_q, \ximat]$ for the evolution under the quench shows that the operators undergo a symplectic transformation $\ximat(t) = \symplectic(t)\ximat(0)$ given by
\begin{equation}
\symplectic(t)
=
\begin{pmatrix}
I & 0 \\
- J t & I
\end{pmatrix}.
\end{equation}
The resulting covariance matrix after the quench is
\begin{equation}
\covariance_{\textnormal{quenched}}
=
\symplectic
\covariance_0
\symplectic^T
=
\frac{1}{2\mu}
\begin{pmatrix}
I & -J t \\
-J t & \mu^2 I + J^2 t^2 \\
\end{pmatrix}.
\label{eq:postquench-covariance}
\end{equation}

We now wish to determine the state of the boundary conditioned on a measurement of the bulk.  To this end, we reorder the overall covariance matrix as
\begin{equation}
\covariance =
\begin{pmatrix}
\Bdy & \Corr \\
\Corr^T & \Bulk
\end{pmatrix},
\end{equation}
where $\Bdy$ and $\Bulk$ are covariance matrices of the boundary and bulk, respectively, and $\Corr$ quantifies the bulk-boundary correlations. The conditional covariance matrix after a partial homodyne measurement in the momentum basis on this state can be calculated as
\begin{equation}
\covariance_{\bdy|\bulk} = \Bdy - \Corr (\Pi_{p}\,\Bulk\,\Pi_{p})^{\mathrm{MP}} \Corr^T
\end{equation}
where $\textnormal{MP}$ is the Moore-Penrose inverse and $\Pi_{p}$ is the projection matrix onto momentum for the bulk subsystem.

The conditional covariance matrix $\covariance_{\bdy|\bulk}$ is independent of the measurement outcomes, with each instance of the protocol preparing a quantum state that is equivalent up to local displacements.  In particular, the conditional mean values $\avg{\ximat}_{\bdy\vert \bulk}$ of the boundary depend on the bulk measurement outcomes $\ximat_\bulk$ as $\avg{\ximat}_{\bdy\vert \bulk}  = \Corr (\Pi_{ p}\,\Bulk\,\Pi_{ p})^{\mathrm{MP}} \ximat_\bulk$~\cite{giedke2002characterization}.  To make the protocol fully deterministic, one may apply a final feedback step consisting of local displacements that shift these post-measurement means to zero.

\textit{Example Geometries}---To illustrate the quench-and-measure protocol, we first consider a quench with $J$ being the adjacency matrix of a graph $G$ whose vertices are the sites. Based on two examples, a hyperbolic disk and a wormhole, we show numerically that the entanglement entropies are well fit by the RT formula in the corresponding geometry.

Consider the graph shown in Figure \ref{fig:setup} inspired by the MERA tensor network~\cite{mera_vidal} and discretizations of the hyperbolic disk. Within AdS/CFT, such a geometry would be associated to the ground state of the boundary CFT, so we compare the numerical entanglement to ground-state CFT predictions. For any CFT defined in one spatial dimension, the entanglement of a subregion of angle $\Delta \theta$ in the ground state on a circle is
\begin{equation}\label{eq:cft-on-circle}
    S_{\textnormal{gs}}(\Delta \theta) = \frac{c}{3} \ln \left( \sin \frac{\Delta \theta}{2}\right) + \epsilon,
\end{equation}
where $c$ is the central charge and $\epsilon$ is a non-universal constant related to the UV cutoff~\cite{Calabrese2004}. Note that $S_{\textnormal{gs}}(2\pi - \Delta \theta)=S_{\textnormal{gs}}(\Delta \theta)$ as required for a translation-invariant pure state. The RT formula applied to AdS$_3$ predicts the same form (reviewed in Supplemental Material~\cite{supp}).

Figure~\ref{fig:setup}(b) compares the numerically obtained entanglement (pink dots) with a fit to the CFT prediction (black dashed line), where $c$ and $\epsilon$ are free parameters. The CFT prediction captures well the overall shape of the data, with only small deviations due to lattice artifacts. \diff{We have also studied two other regular graph discretizations of the disk with comparable results (see Supplemental Material~\cite{supp}).} The fit value of $c$ scales with $\ln \frac{1}{\mu} $ when $\mu$ is small, and the shape of the entropy curve depends to some extent on the magnitude of $\mu$ (see Supplemental Material~\cite{supp}). 

Full-featured holographic duals also have R{\'e}nyi entropies with non-trivial dependence on the R{\'e}nyi parameter. The R{\'e}nyi entropy of a density matrix $\rho$ is defined as $S_\alpha(\rho)= \frac{1}{1-\alpha}\ln \text{tr}(\rho^\alpha)$, and it can also be computed from the covariance matrix~\cite{supp}. For our setting with the disk graph, the results are shown in Figure~\ref{fig:setup}(c) for different values of the squeezing $\mu$ (colored curves).  We see a non-trivial dependence on the R{\'e}nyi parameter, in contrast to simple tensor-network models, and by tuning the squeezing we can come close to the CFT result $(S_\alpha / S)^\mathrm{CFT} = (1+1/\alpha)/2$ (black dashed line)~\cite{Calabrese_2009}.

To demonstrate the power of the framework beyond the hyperbolic disk case, we also consider a wormhole geometry in which we take two copies of the graph for the hyperbolic disk, remove identical regions of the interior of each graph, and then glue the graphs together by adding corresponding links in the adjacency matrix. The resulting graph is shown in Figure~\ref{fig:wormhole}(a). This wormhole geometry mimics a spatial slice of an eternal AdS black hole, and would be dual to the thermofield double state under AdS/CFT~\cite{Maldacena_2003}. We consider the entanglement entropy for two different types of regions: either an interval on one side of the wormhole [Fig.~\ref{fig:wormhole}(b)] or a pair of corresponding intervals on both sides [Fig.~\ref{fig:wormhole}(c)].

\begin{figure}
    \centering
    \includegraphics[width=\linewidth]{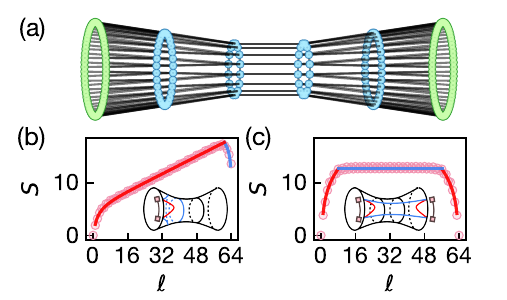}
    \caption{(a) A wormhole graph, formed from gluing together portions of two hyperbolic disk graphs. In this case, each boundary has $L=64$ sites. (b) One-sided entanglement entropy of the wormhole for $\mu = 0.2$, with the
    fit yielding inverse temperature $\beta=2.9(1)\times10^1$, central charge $c=5.8(2)$, and offset $\epsilon=2.1(1)$. The fit is shown in red and blue, which correspond to two different candidate curves for the minimal surface (see inset). (c) Two-sided entanglement entropy of the wormhole for $\mu = 0.2$. The approximate condition used to find the crossover between the red and blue candidate two-sided entanglement entropies for the model curve is given in~\cite{supp}.}
    \label{fig:wormhole}
\end{figure}

For both the one-sided and two-sided regions, the entanglement entropy is well fit by minimal curves obtained from the (2+1)D eternal AdS black hole geometry~\cite{PhysRevLett.69.1849} (see Supplemental Material for a review of the relevant geodesics~\cite{supp}). In both cases, there is a competition between two different candidate minimal surfaces, shown in red and blue on the wormhole schematics in Fig.~\ref{fig:wormhole}(b-c). The RT formula predicts a sharp change in the entanglement entropy when these two candidates swap dominance, and the calculated entanglement entropy (pink circles) shows a corresponding rapid crossover from one candidate to another (red and blue fit curves).  \diff{We provide evidence in the Supplemental Material that holographic inequalities such as monogamy of mutual information are satisfied by these states~\cite{supp}.}

\textit{Power laws from decoration}---These two examples of the quench-and-measure protocol yield states with entanglement entropies that are well approximated by minimal curves in a hyperbolic disk and wormhole geometry. They also have non-trivial R{\'e}nyi entropies which can match CFT predictions. To also realize clean power-law correlations, it is useful to introduce an additional structure to the quench Hamiltonian which we refer to as decoration. An example of a decorated graph is illustrated in Figure \ref{fig:decorated-graph}(a). It is constructed by first choosing a bulk graph (blue nodes) and choosing which bulk nodes are adjacent to boundary nodes (green nodes). The edges of the bulk graph are then split into two by adding intervening bulk nodes (yellow nodes). For each pair of split edges, one is assigned $+1$ in the $J$ matrix defining the quench Hamiltonian and the other is assigned $-1$.

The decoration procedure can be motivated by considering just three modes, labeled $1$, $2$, and $3$. If all modes are initialized in an unentangled state, quenched with a $J_{12} x_1 x_2 + J_{23} x_2 x_3$ interaction, and mode $2$ is subject to a momentum measurement, then the resulting correlations $\avg{x_1 x_3}$ have a sign given by $- \text{sgn}(J_{12} J_{23})$. Hence, if $J_{12}$ and $J_{23}$ have opposite signs, then the resulting post-measurement correlation is positive. The split nodes in the decorated graph thus ensure a non-alternating pattern of correlations~\cite{supp}.

As an example, we consider the decorated graph in Fig.~\ref{fig:decorated-graph}, derived from the hyperbolic disk graph in Fig.~\ref{fig:setup}. We examine the state prepared by the quench-and-measure protocol on this graph for strong initial squeezing, $\mu = 0.05$.  In this regime, both the CFT ground-state entanglement entropy formula and power-law decay for the position-position two-point function are approximately satisfied, the latter exhibiting a scaling dimension of $\Delta_x \approx 1$. The scaling dimension of an operator $\phi$ 
is extracted from covariance by the relationship $\mathrm{Cov}(\phi_i, \phi_j) \propto  \sin(\pi |i - j|/L)^{- 2\Delta_\phi}$. In this strong squeezing regime, the momentum-momentum
correlations are approximately constant with distance, corresponding to a scaling dimension $\Delta_p \approx 0$. The central charge fit to the CFT entanglement entropy saturates at $c=1$ (see Supplemental Material~\cite{supp}), which is the same as for a (1+1)D free scalar CFT.

\diff{To provide further evidence for the CFT nature of the decorated state, one can appeal to the strong constraints imposed by conformal symmetry~\cite{lin2023conformal,li2025systematicsearchconformalfield}. From symmetry alone, one can deduce the form of the reduced density matrix of an interval and use this to test if a state is a CFT ground state. In the SM we apply this test, known as the vector fixed point equation, to the state prepared from the decorated MERA graph and find that it approximately satisfies the equation with a residual that vanishes with system size. Hence, we have strong evidence that the decorated MERA state approaches the ground state of the free scalar CFT in the continuum limit.}

\begin{figure}
    \centering
    \includegraphics[width=\linewidth]{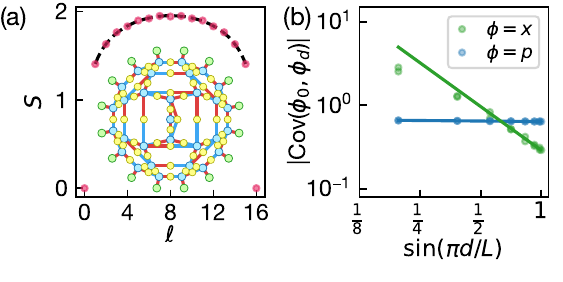}
    \caption{(a) Entanglement entropies for the boundary of a depth-5 decorated graph for $\mu = 0.05$, along with the CFT fit (dashed black line) yielding $c=1.004(9)$ and vertical offset $\epsilon=1.954(2)$. The inset shows the decorated graph, where green are boundary nodes, blue are the original bulk nodes, and yellow are additional bulk nodes. The color of the edge indicates the sign of the interaction (red for positive, blue for negative). (b) Position (green) and momentum (blue) covariances of a pair of boundary sites vs distance $d$, where the boundary circumference is $L=16$. Lines show power-law fits at large $\sin(\pi d / L)$, yielding scaling dimensions $\Delta_x = 0.87(4)$ and $\Delta_p = 1.03(3)\times10^{-2}$.
    }
    \label{fig:decorated-graph}
\end{figure}

\textit{Bulk reconstruction}---Having demonstrated entropy results consistent with the RT formula in a variety of examples, a natural next question is whether the putative RT surface can be directly detected. To study this, we consider a problem of ``bulk reconstruction'' defined as follows. Consider, for example, the undecorated protocol for a graph $G$. Choose any bulk site and modify the graph by adding one boundary site which connects to the chosen bulk site. We refer to this new boundary site as the probe and emphasize that, although it connects directly to the bulk, it is not measured. Measure the bulk sites as before to construct a pure boundary state, now including the additional probe site. The physics of the probe is conveniently discussed in terms of the mutual information, $I(A:B) \equiv S(A) + S(B) - S(A \cup B)$. Because the probe is entangled with the rest of the boundary and the total state is pure, the mutual information between the probe $b$ and the rest of the boundary is
\begin{equation}
    I(\bdy:b) = 2 S(b).
\end{equation}
More generally, if $A$ is a subset of the rest of the boundary, then the entanglement with the probe can be recovered just from $A$ if the mutual information satisfies $I(A:b) = 2 S(b)$~\cite{Xu_2024}.

By fixing a boundary subset $A$ and varying the location of the probe, we can use the normalized mutual information, $I(A:b)/S(b)$, to map out the degree to which a bulk site can be reconstructed from the given boundary region. The connection to the RT surface arises because, in a holographic system obeying the RT formula, one can show that the normalized mutual information is $2$ when the probe is connected to a bulk site inside the RT surface of the given boundary region and zero when it is connected to a bulk site outside. Hence, the boundary between the reconstructible and non-reconstructible regions, defined as $I(A:b)/S(b) > 1$ and $< 1$ respectively, should track the RT surface. 

As a proof of principle, we demonstrate this diagnostic in the MERA/hyperbolic disk case. The four panels in Figure \ref{fig:reconstruction} show different boundary regions: a single interval (top) or a pair of intervals of different sizes (bottom). The pink highlighted boundary sites indicate the boundary region of interest, with the other boundary sites being green. The bulk sites are then colored ranging from pink to black indicating the size of the corresponding normalized mutual information. The dark pink solid line in the bulk indicates the RT prediction, in this case segments of circles on the Poincar{\'e} disk model of hyperbolic space. We see reasonably good agreement with the holographic prediction, with bulk sites deep inside the RT surface having high mutual information and bulk sites far outside having very low mutual information. However, the boundary is not completely sharp, suggesting a thickening or fluctuation of the RT surface that merits future investigation.

\begin{figure}
    \centering
    \includegraphics[width=0.9\linewidth]{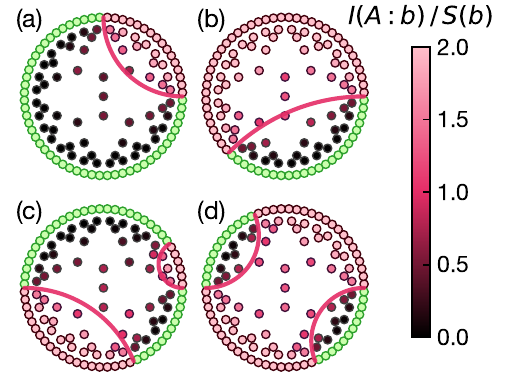}
    \caption{Approximate reconstruction of the entanglement wedge through mutual information. The color of each bulk node shows the normalized mutual information $I(A:b) / S(b)$ between a subset of boundary nodes (light pink nodes on the ring) and an unmeasured probe $b$ connected to the bulk node. Note that $2S(b)$ bounds $I(A:b)$. This undecorated construction uses $\mu=1$, which leads to a sharper surface compared to stronger initial squeezing. The expected minimal surface is drawn in dark pink. The boundary regions $R$ are (a) a small connected region (b) a large connected region (c) two disconnected regions with an expected disconnected entanglement wedge, and (d) two disconnected regions with an expected connected entanglement wedge.}
    \label{fig:reconstruction}
\end{figure}

\textit{Outlook}---We introduced a Gaussian quench-and-measure protocol and gave ample numerical evidence that it prepares states with holographic entanglement dictated by a bulk geometry of our choice. Our approach is inspired by tensor network models, and it naturally includes additional features such as non-trivial R{\'e}nyi entropies and the possibility of power-law correlations.  As the states output by our protocol are Gaussian, their entanglement properties can be efficiently reconstructed from measured correlations~\cite{tajik2023verification}. Moreover, by leveraging measurement rather than relying on dynamical spreading of correlations, the protocol generates long-range entanglement in $O(1)$ quench time. Hence, our proposal provides an efficient framework to prepare and probe holographic states.

\diff{It is surprising that one can prepare states with holographic entanglement using only Gaussian operations.  While admitting states that satisfy the RT formula, the Gaussian setting cannot capture all features of a full-fledged model of quantum gravity.  Our protocol nevertheless offers a simplified setting for studying a key aspect of holography in isolation, as well as a potential starting point for generalizations incorporating, e.g., non-Gaussianity or more complex local degrees of freedom.}

Our proposal is amenable to instantiation on several experimental platforms.  At the heart of our protocol is the quench that prepares a continuous-variable graph state.  Such states have been generated in photonic platforms~\cite{chen2014experimental,asavanant2019generation,larsen2021deterministic} and in arrays of atomic ensembles with photon-mediated interactions~\cite{cooper2024graph,periwal2021programmable}, both providing flexible control over the graph.  A further possibility is to leverage superconducting circuits featuring hyperbolic lattices of coplanar waveguide resonators~\cite{kollar2019hyperbolic}.  Our protocol should also generalize to stabilizer states of qubits, opening the door to implementation in quantum computing platforms including reconfigurable arrays of Rydberg atoms~\cite{bluvstein2022quantum} or trapped ions~\cite{miao2024probing,iqbal2024topological}.

Beyond their fundamental interest, the long-range entangled states generated by our protocol may serve as resources for sensing or computation.  The scale-invariant correlations in the Gaussian setting could enable entanglement-enhanced imaging of spatially structured fields.  Generalizations to stabilizer states of qubits, or incorporation of non-Gaussianity, may enable novel error correcting codes~\cite{Breuckmann_2016} and bring in new features of holography~\cite{cao2025gravitationalbackreactionmagical}. An important question for future work is whether our protocol can accommodate a degree of non-Gaussianity while keeping post-selection manageable.  Future work may explore how experimental imperfections impact the state preparation and to what extent entanglement in such imperfectly prepared states can be described by a quantum extremal surface formula~\cite{Engelhardt:2014gca}.

Our protocol also raises broader questions for future theoretical work benefiting from the tractability of the Gaussian setting. Prospects include seeking a deeper analytical understanding of the entanglement structure, a study of entropy inequalities and the entropy cone~\cite{Bao_2015}, \diff{measures of multi-party entanglement~\cite{Akers_2022,PhysRevLett.125.241602},} perturbations and the problem of metric reconstruction~\cite{Cao_2020}, probes of complexity~\cite{Baiguera_2026}, and connections between wormholes and random Gaussian states~\cite{Magan_2025}. \diff{Among these, we are actively working on the entropy cone question, with a preliminary study of the tripartite mutual information for the undecorated disk in the SM~\cite{supp}.}

\begin{acknowledgments}
We thank Eugene Demler, Soonwon Choi, Nadie Li, Brian Khor, Allison Powell, Merrick Ho, and Ocean Zhou for discussions. We gratefully acknowledge support from the Heising-Simons Foundation under grants 2024-4849 (B.S.) and 2024-4850 (J.~J. and M.S.-S.), the National Science Foundation under award number 2409479 (M.~S.-S.), and the DOE Office of Science, Office of High Energy Physics under award DE-SC0025934 (L.~G.).
\end{acknowledgments}

\endgroup

\bibliography{references}

\clearpage
\onecolumngrid
\SMCombinedTitle{Supplementary Information for ``Building Holographic Entanglement by Measurement''}

% begin supplemental numbering
\setcounter{figure}{0}
\setcounter{equation}{0}
\setcounter{section}{0}
\setcounter{subsection}{0}

% Sections: I, II, III...
\renewcommand{\thesection}{\Roman{section}}

% Subsections: 1, 2, ...
\renewcommand{\thesubsection}{\arabic{subsection}}

% Prepend "S" to figure/equation numbers
\renewcommand{\thefigure}{S\arabic{figure}}
\renewcommand{\theHfigure}{S\arabic{figure}}
\renewcommand{\theequation}{S.\arabic{equation}}
\renewcommand{\theHequation}{S.\arabic{equation}} % add this

% Hyperref anchors for sectioning (must be unique)
\renewcommand{\theHsection}{SM.\Roman{section}}
\renewcommand{\theHsubsection}{SM.\Roman{section}.\arabic{subsection}}

% This is so we number the sections even with RevTeX
\setcounter{secnumdepth}{2}

\StartSupplementTOCEntries

\supplementtableofcontents

\vspace{2em}

In this supplement we expand upon the background theory and constructions used for
the quench-and-measure protocol in the main text.
In Sec.~\ref{sec:review} we review the continuous geometries that we will discretize with graph constructions. In Sec.~\ref{sec:discretizing} we motivate the graphs we use for discretizations.
In Sec.~\ref{sec:correlations-and-entanglement}, we present additional numerical data showing how correlations and entanglement entropies depend on the graph parameters and
the specific construction (undecorated vs. decorated) chosen,
and compare correlations to mutual information.
In Sec.~\ref{sec:gaussian-states} we provide a self-contained review of the calculation of entanglement properties of Gaussian states from the covariance matrix.
Finally, in Sec~\ref{sec:vfpe} we discuss the vector fixed point equation as a measure of how
the decorated construction can converge to a CFT for large graphs.

\section{\texorpdfstring
  {Review of $(2+1)$d geometries appearing in AdS$_3$/CFT$_2$}
  {Review of (2+1)d geometries appearing in AdS3/CFT2}}\label{sec:review}

In the main text we consider two classes of geometries to demonstrate the quench-and-measure protocol. The first of these is the hyperbolic disk, which we think of as a spatial slice of AdS$_3$. The second is the wormhole geometry, which we think of as the spatial slice of two-sided BTZ black hole geometry. In AdS/CFT, the first of these geometries would correspond to the ground state of the dual CFT. The second would correspond to a thermal state of the CFT. Here we review these geometries and show how the RT surfaces referenced in the main text may be obtained. 

The key technical fact is that all the geometries of interest can be presented using an embedding space approach. We embed each $(2+1)$D geometry into a $(2+2)$D space with coordinates $(T_1,T_2,X_1,X_2)$ and metric
\begin{equation}
    ds^2_4 = - dT_1^2 - dT^2_2 + dX_1^2 + dX^2_2. \label{eq:4d_metric}
\end{equation}
This is a flat spacetime with two time-like and two space-like directions. The embedding takes the form of a hyperboloid
\begin{equation}
    -T_1^2 - T_2^2 + X_1^2 + X^2_2 = - \lads^2.
\end{equation}
Here $\lads$ is the curvature radius also known as the AdS radius.
This presentation is convenient because it gives a simple formula for the geodesic distance between two points in the $(2+1)$D space. If $P$ and $Q$ are points on the hyperboloid, then the $(2+1)$D distance obeys 
\begin{equation}
    \cosh \frac{\db }{\lads} = - \frac{P\cdot Q}{\lads^2}\label{eq:geodesic}
\end{equation}
where $\cdot$ denotes the inner product in the $(2+2)$D metric \eqref{eq:4d_metric}.

This framework is convenient for our purposes because, in AdS/CFT applied to a $(2+1)$D bulk, the minimal surface that computes the entanglement reduces to a minimal curve, i.e. a geodesic. So by fixing the boundary region of interest using the points $P$ and $Q$, we can read off the relevant geodesic lengths from \eqref{eq:geodesic}. Then the entanglement entropy of the boundary region is obtained from the RT formula, 
\begin{equation}
    S = \frac{\db}{4G}.
\end{equation}
The Brown-Henneaux relation for the central charge, $c = \frac{3\lads}{2G}$, allows us to convert from bulk units to the central charge characterizing the CFT. We also note that, as discussed in the main text, it is often necessary to perform an outer overall minimization over different candidate RT surfaces built from local geodesics.

\subsection{\texorpdfstring
  {AdS$_3$}
  {AdS3}}
We first discuss the case of AdS, which corresponds to the CFT ground state. The coordinates in $(2+1)$D are  $(r,t,\theta)$ and the embedding map is
\begin{align}
    & T_1 = \sqrt{\lads^2 +r^2} \sin \frac{t}{\lads}, T_2 = \sqrt{\lads^2 +r^2} \cos \frac{t}{\lads}, \nonumber \\ & X_1 = r \cos \theta, X_2 = r \sin \theta,
\end{align}
which leads to the induced metric
\begin{equation}
    ds_3^2 = - \left( 1 + \frac{r^2}{\lads^2}\right) dt^2 + \frac{1}{1+\frac{r^2}{\lads^2}} dr^2 + r^2 d\theta^2.
\end{equation}
Introduce a large $r$ cutoff at $r=r_c$ and consider a cutoff-anchored geodesic from $(0,r_c,0)$ to $(0,r_c,\Delta \theta)$. The corresponding $(2+2)$D points are 
\begin{align}
    & P = (0,\sqrt{\lads^2 + r_c^2},r_c ,0), \nonumber \\ & Q = (0,\sqrt{\lads^2 + r_c^2},r_c \cos \Delta \theta, r_c \sin \Delta \theta).
\end{align}
Thus the distance obeys
\begin{equation}
    \cosh \frac{\db}{\lads} = 1 + \frac{r_c^2}{\lads^2} (1 -  \cos \Delta \theta).
\end{equation}
Using the fact that the cutoff $r_c$ is large, we obtain the simplified form
\begin{equation}
    \db \approx \lads \ln \left( 4 \frac{r_c^2}{\lads^2} \sin^2 \frac{\Delta \theta}{2} \right).
\end{equation}

The entropy is
\begin{equation}
    S = \frac{\db}{4G} = \frac{\lads}{2G} \ln \left( \frac{2 r_c}{\lads} \sin \frac{\Delta \theta}{2} \right),
    \label{eq:ads_ent_cutoff}
\end{equation}
or, using the Brown-Henneaux relation,
\begin{equation}
    S = \frac{c}{3} \ln \left( \sin \frac{\Delta \theta}{2}\right) + \text{(constant)}. \label{eq:ent_gs}
\end{equation}
This formula is invariant under the replacement $\Delta \theta \to 2 \pi - \Delta \theta$ as required by the purity of the ground state.

\subsection{BTZ Black Hole}
We next discuss the case of the non-rotating BTZ black hole, which corresponds to a CFT thermal state. Specifically, we consider a two-sided black hole that corresponds to a purification of the mixed thermal state into a thermofield double state (TFD). The coordinates in $(2+1)$D are again $(r,t,\theta)$ and the embedding map is 
\begin{align}
    & T_1 =  \frac{\lads \sqrt{r^2 - r_+^2}}{r_+} \sinh \frac{r_+ t}{\lads^2}, T_2 =  \frac{\lads r}{r_+} \cosh \frac{r_+ \theta}{\lads}, \nonumber \\ & X_1 = \frac{\lads \sqrt{r^2 - r_+^2}}{r_+}  \cosh \frac{r_+ t}{\lads^2}, X_2 =  \frac{\lads r}{r_+} \sinh \frac{r_+ \theta}{\lads}, 
\end{align}
with $r_+$ a constant. The induced $(2+1)$D metric is
\begin{equation}
    ds^2_3 = - \frac{r^2 - r_+^2}{\lads^2} dt^2 + \frac{\lads^2}{r^2 - r_+^2} dr^2 + r^2 d\theta^2
\end{equation}
which has a horizon at $r=r_+$. The location of the horizon fixes the thermodynamic properties of the solution, including its temperature
\begin{equation}
    T = \frac{r_+}{2\pi \lads^2}
\end{equation}
and thermal entropy 
\begin{equation}
    S_{th} = \frac{2 \pi r_+}{4G} = \frac{\pi^2 T \lads^2}{G} = \frac{2 \pi^2 c \lads T}{3 }.
\end{equation}

One can identify the physical length of the circle on which the CFT lives with the proper distance at the cutoff surface,
\begin{equation}
    L_{\text{CFT}} = 2 \pi r_c.
\end{equation}
Setting $x_{\text{CFT}} = r_c \theta$ and $t_{\text{CFT}} = \frac{r_c}{\ell} t$, the induced metric on the cutoff surface is
\begin{equation}
    ds^2_{\text{CFT}} = - dt_{\text{CFT}}^2 + dx_{\text{CFT}}^2.
\end{equation}
In this time coordinate, the CFT temperature is
\begin{equation}
T_{\text{CFT}} = \frac{\lads}{r_c} T.    
\end{equation}
The thermal entropy is then 
\begin{equation}
    S_{th} = \frac{\pi c L_{\text{CFT}} T_{\text{CFT}}}{3},
\end{equation}
which is the standard formula for $(1+1)D$ CFT.

We now consider the one-sided and two-sided entropies, starting with the one-sided case. For the same cutoff anchored boundary points as in the AdS$_3$ case, the embedding space points are now
\begin{align}
   & P = (0, \frac{\lads r_c}{r_+} , \frac{\lads r_c}{r_+} , 0), \nonumber \\ & Q = (0, \frac{\lads r_c}{r_+} \cosh \frac{r_+ \Delta \theta}{\lads}, \frac{\lads r_c}{r_+}, \frac{\lads r_c}{r_+} \sinh \frac{r_+ \Delta \theta}{\lads} ).
\end{align}
With the same large $r_c$ simplification, the distance is
\begin{equation}
    \exp \frac{\db}{\lads} = \frac{4 r_c^2}{r_+^2} \sinh^2 \frac{r_+ \Delta \theta}{2\lads} .
\end{equation}

Taking this geodesic as a candidate for the RT surface, we get a candidate for the entropy, 
\begin{equation}
    S^{(1)} = \frac{\db}{4G} = \frac{c}{3} \ln \left(\frac{2 r_c}{r_+} \sinh \frac{r_+ \Delta \theta}{2\lads} \right). \label{eq:ent_1d_opt1}
\end{equation}
An alternative geodesic is provided by the union of the horizon and the analog of the above geodesic but for the complementary boundary region. The corresponding entropy candidate is 
\begin{equation}
    S^{(2)} = S_{th} +\frac{c}{3} \ln \left(\frac{2 r_c}{r_+} \sinh \frac{r_+ (2 \pi - \Delta \theta)}{2 \lads} \right).  \label{eq:ent_1d_opt2}
\end{equation}
The general formula for the one-sided entropy predicted by RT is thus
\begin{equation}
    S = \min(S^{(1)},S^{(2)}). \label{eq:ent_1side}
\end{equation}

In the two-sided case, one candidate RT surface is a union of two one-sided surfaces, one for the boundary region in the left asymptotic region and one for the boundary region in the right asymptotic region. Another option is a surface that connects the boundary region end points in the left asymptotic region to the corresponding endpoints in the right asymptotic region. Each of these two geodesics is itself composed of a geodesic from $r=r_c$ to $r=r_+$ at constant $t$ and $\theta$. The points in this case are
\begin{align}
   & P = (0, \frac{\lads r_c}{r_+} , \frac{\lads r_c}{r_+} , 0), \nonumber \\ & Q = (0, \lads, 0,0 ).
\end{align}
This partial distance obeys 
\begin{equation}
    \exp \frac{\db'}{\lads} = \frac{2 r_c}{r_+},
\end{equation}
so the full distance is
\begin{equation}
    \db = 4 \db' = 4 \lads \ln \frac{2 r_c}{r_+}.
\end{equation}

We may obtain the same formula by identifying the second asymptotic region with $t \to t + i \beta/2$. The geodesic length from the left boundary endpoint $(0,r_c,0)$ to the right boundary endpoint $(i \beta/2, r_c , 0)$ can be found from the points
\begin{align}
   & P = (0, \frac{\lads r_c}{r_+} , \frac{\lads r_c}{r_+} , 0), \nonumber \\ & Q = (0, \frac{\lads r_c}{r_+} , - \frac{\lads r_c}{r_+}, 0 ).
\end{align}
The corresponding distance obeys
\begin{equation}
    \exp \frac{\db''}{\lads} = 4 \frac{ r_c^2}{r_+^2},
\end{equation}
and the full distance is now twice $\db''$ giving again $\db = 4 \lads \ln \frac{2 r_c}{r_+}$.

The corresponding candidate two-sided entropy is 
\begin{equation}
    S^{(\text{two-sided, alt})} = \frac{2 c}{3} \ln \frac{2 r_c}{r_+}.
\end{equation}
Comparing to $2 S^{(1)}$ from the one-sided calculation, these two surfaces exchange relevance when $\Delta \theta$ satisfies
\begin{equation}
\sinh \frac{r_+ \Delta \theta}{2 \lads} = 1,
\end{equation}
which is the condition used to estimate the location
of the crossover between the two entanglement entropy curves in Fig.~\MainRef{fig:wormhole}(c) of the main text. Note that this calculation assumes no interior to the wormhole; if there is an interior then the two-sided alternate lengths will be longer and hence the transition to the plateau will occur at larger interval size.

\section{Discretizing bulk geometries with graphs}\label{sec:discretizing}

Our protocol requires the discretization of a spatial slice of a continuous bulk geometry into a
representative graph. The chosen graph should approximate the bulk geometry in the following sense:
if each node in the graph is identified with a position in the bulk, edges should only exist between nodes nearby in proper distance. All edges in the graph should have approximately the same associated proper distance,
which justifies use of an unweighted graph to represent the spatial slice of the bulk geometry.

\subsection{Discretizing the hyperbolic disk: MERA graph}

To discretize the hyperbolic disk, we use a graph $G$ inspired by the MERA tensor network~\mcite{vidal2007entanglement,swingle2012entanglement}.  This graph is formed by combining a tree graph of branching ratio $b=2$ with additional edges that connect the nodes within each level of the tree into a ring, as shown in Fig.~1(a) of the main text.  For a tree of total depth $\depth_b$, the boundary has $L = 2^{\depth_b}$ nodes, while the bulk has $N_\bulk = 2^{\depth_b} - 1$ nodes, yielding in total $N_\bulk + L = 2^{\depth_b+1}-1$ nodes. 

We can now associate each bulk node with an approximate position inside the continuous geometry of the AdS$_3$ spatial slice. We consider $r \gg \lads$ and define
\begin{equation}
    d\scriptr = \lads \frac{dr}{r}.
\end{equation}
With the constraint that the proper distance between adjacent levels of the tree should equal the proper
distance between adjacent nodes within a level of $2^{\depth}$ nodes at depth $\depth$,
we find
\begin{equation}
r = \frac{\lads \ln 2}{2\pi} e^{\scriptr/\lads}
\end{equation}
such that the approximate spatial metric can be written as
\begin{equation}
ds^2 = d\scriptr^2 +{ \left(\frac{\lads \ln2}{2\pi} \right)}^2 e^{2\scriptr / \lads}d\theta^2.
\end{equation}
Here $d_{\mathrm{adj}} = \lads \ln 2$ is the approximate proper distance between adjacent nodes. 
In these coordinates, bulk nodes in a layer of depth $\depth$ are
placed at $\scriptr = R\,d_{\mathrm{adj}}$ and are spaced within a layer by $\Delta\theta = 2\pi/L = 2\pi/2^n$.

\subsection{Discretizing the wormhole}

As discussed in Sec.~\ref{sec:review}, by ``wormhole'' geometry we refer to the spatial slice of the two-sided BTZ black hole. For large $r$, the one-sided BTZ metric approaches AdS$_3$. To form the two-sided BTZ solution, two one-sided BTZ geometries are glued together at $r_+$. This suggests that as an approximation one can discretize the two-sided BTZ solution by gluing together two graphs that each approximate a large-$r$ region of AdS$_3$.

To accomplish this, we take two copies of the hyperbolic disk graph, and remove nodes with depth less than some integer $\depth_+$. The nodes at depth $\depth_+$ in each graph can either be identified (in the case of no wormhole interior) or connected through a cylindrical mesh to represent the wormhole interior. In Fig.~2(c) in the main text, we show the case where an edge is drawn between each node at depth $\depth_+$ and equal angular position in the two copies of the hyperbolic disk graph.

\subsection{Decoration}\label{sec:decoration}

\begin{figure}[!htbp]
    \centering
    \includegraphics[width=0.75\linewidth]{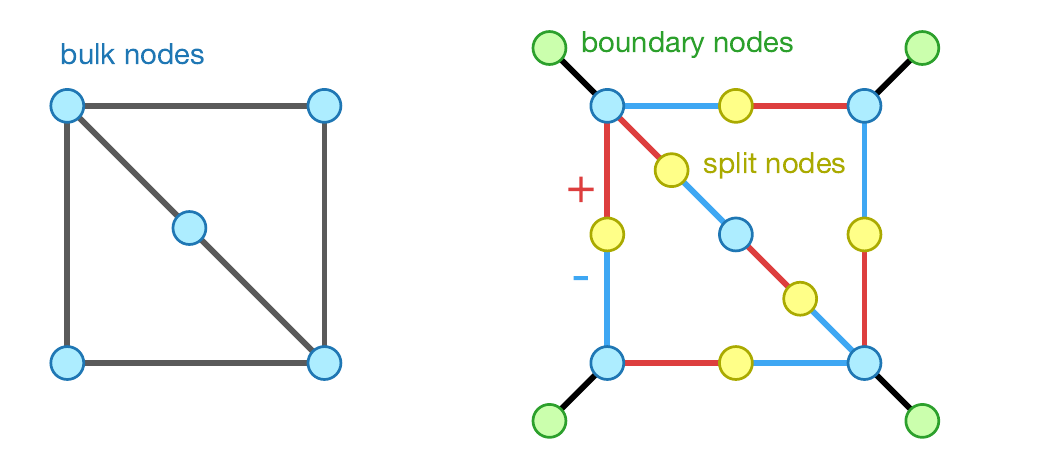}
    \caption{An illustration of the decoration construction. \textbf{Left:} The initial bulk graph (blue nodes) used to form the decorated graph. In this example, the outermost four sites are chosen to be adjacent to the boundary. \textbf{Right:} The added boundary nodes (green) are connected to the pre-selected adjacent bulk nodes. All bulk-to-bulk edges are also divided with split nodes (yellow), with the adjacent pair of links assigned $+$ or $-$ in the coupling matrix.}
    \label{fig:sm_decorate}
\end{figure}

Decoration, which is illustrated in Fig.~\ref{fig:sm_decorate}, is a procedure that alters the graph used to discretize a spatial geometry. Take a graph $G=(V,E)$ with vertices $V$ and unweighted edges $E$. In this case, all of the original vertices should be thought of as bulk vertices. Given $G$ and a subset $V_{O} \subset V$,
we define a decorated graph $G_D = (V_D,E_D)$ with vertices $V_D$ and weighted edges $E_D$ in the following manner.
First, we subdivide the graph, splitting each edge $e\in E$ into two edges $e^+ \in E^+$ and $e^- \in E^-$ by inserting vertices. These inserted vertices, which we call ``split'' nodes, form the subset $V_S \subset V_D$. Edges in $E^+$ have weight $+1$ and edges in $E^-$ have weight $-1$. Finally,
for every vertex $v_O \in V_O$, a boundary vertex $v_\bdy \in V_\bdy$ is added to the graph and
connected to $v_O$ via an edge with weight $+1$. The vertex set of $G_D$ is thus
\begin{equation}
    V_D = V \cup V_S \cup V_\bdy
\end{equation}
and it contains $|V| + |E| + |V_{\bdy}|$ vertices in total. The edge set $E_D$ contains two edges for each original edge, from the splitting procedure, and a new edge for each boundary vertex. It contains $2|E| + |V_{\bdy}|$ edges in total.

The quench Hamiltonian associated with the decorated graph $G_D$ follows its (weighted) adjacency matrix $A_D$. The quench Hamiltonian has the form
\begin{equation}
    H_q = \frac{1}{2} \sum_{\langle ij \rangle\in E_D} J_{ij} x_i x_{j},
\end{equation}
where $J_{ij} = A_{D, ij}$. During the measurement process of the quench-and-measure protocol,
both $V$ and $V_S$ are measured.

The decorated graph yields much cleaner power laws of correlations than the undecorated case,
because contributions to the covariance from different paths starting and ending on the same boundary anchors constructively add, as described in the main text.
The decorated construction was also motivated by attempts to embed the Laplacian $L = D - A$ of a graph $G$ into the covariance matrix produced after the quench.
The appearance of the Laplacian in the post-quench, pre-measurement covariance matrix of the decorated procedure can be noted by
examining $A_D^2$, which appears in the momentum-momentum block of the covariance matrix (cf. Eq.~\MainEqref{eq:postquench-covariance} of the main text).
The values of $A_D^2$ are a weighted sum of length-2 paths,
such that due to the form of decoration, the Laplacian $L$ of the original graph contributes to the bulk node block of $A_D^2$. We were motivated to engineer the appearance of the bulk Laplacian because of a basic element of the AdS/CFT dictionary which relates a boundary operator of scaling dimension $\Delta$ to a bulk field of mass $m_{\text{cont}} \sim \Delta/\lads$. We speculated that scale-invariant correlations could be obtained by mimicking the appearance of a massive free field in the bulk. Since the lattice Hamiltonian of a continuum massive free field of mass $m_{\text{cont}}$ is set by $L + m_{\text{cont}}^2$, the Laplacian was a natural target. In practice, calculations showed that decoration produces cleaner entanglement entropy whose fit central charge saturates to near $c=1$ at strong squeezing (see the lower panel (c) of Fig.~\ref{fig:entanglement_entropy_comparison}). This is suggestive of a free scalar CFT on the boundary, a suggestion which we revisit below.

\diff{
For a summary comparing which holographic or CFT conditions are satisfied by the undecorated and decorated constructions, please see Table~\ref{table:condition-satisfied}. The evidence supporting this breakdown is given in Sec.~\ref{sec:correlations-and-entanglement} and Sec.~\ref{sec:vfpe}.
}

\begin{table*}[!htbp]
\caption{\label{tab:results}\diff{Conditions satisfied or violated for different constructions.}}
\label{table:condition-satisfied}
\begin{ruledtabular}
\begin{tabular}{lccc}
Condition & Satisfied when & Undecorated Graph & Decorated Graph \\
\hline
Ryu-Takayanagi formula & state is holographic & approximately satisfied & only single-interval \\
Monogamy of mutual information (MMI) & state is holographic & approximately satisfied & violated \\
Vector fixed point equation (VFPE)& state is CFT ground state & violated & approximately satisfied\\
\end{tabular}
\end{ruledtabular}
\end{table*}

\subsection{\diff{Truncations of regular graphs}}\label{sec:regular-graphs}

In the main text, the MERA graph was used to discretize the hyperbolic disk. However, other discretizing graphs
can be used to obtain similar entropy results as long as they satisfy the approximate conditions described in the beginning of Sec.~\ref{sec:discretizing}.

\begin{figure}[!htbp]
    \centering
    \includegraphics[width=0.75\linewidth]{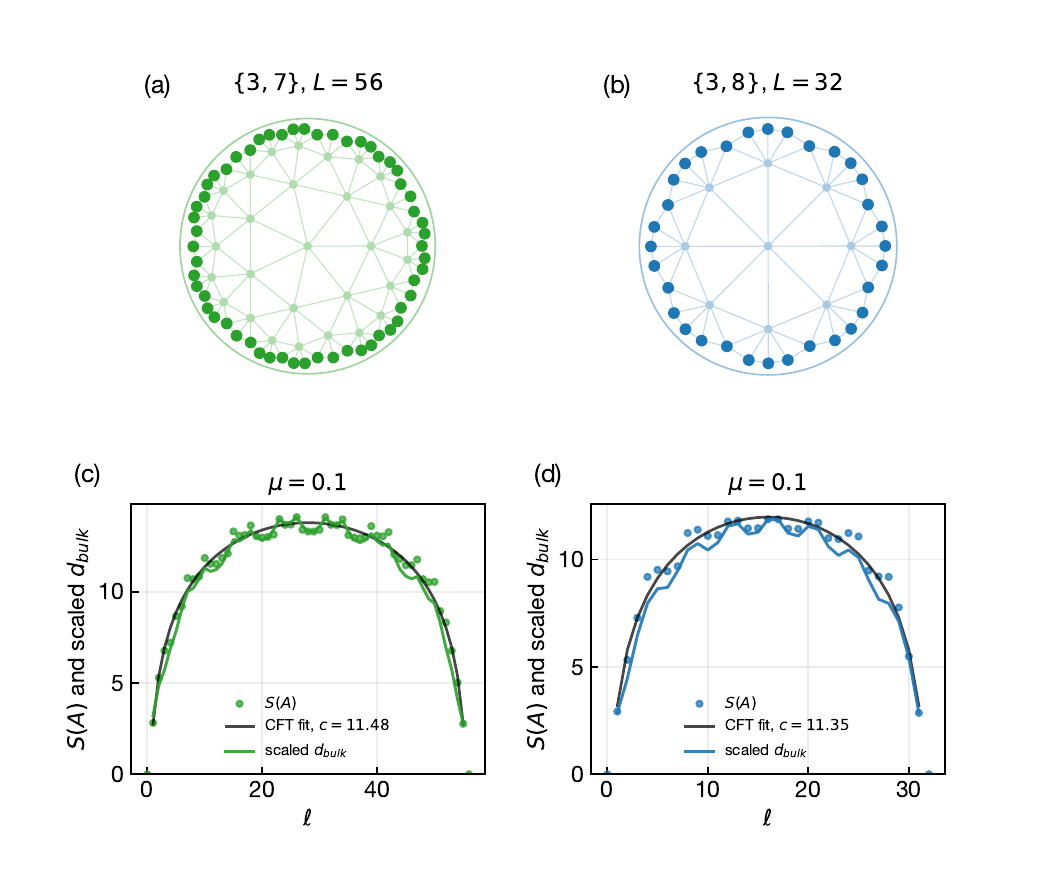}
    \caption{\diff{Regular hyperbolic graphs in the undecorated case using $\mu=0.1$ for (a) a $\{3,7\}$ graph and (b) a $\{3,8\}$ graph. A scaled $d_{bulk}$, which is estimated through the positions of points on the hyperbolic plane, is plotted alongside
    entropy of boundary regions in (c) the $\{3, 7\}$ graph and (d) the $\{3, 8\}$ graph.
    }}
    \label{fig:regular-graphs}
\end{figure}

For example, one may consider truncations of regular graphs that tile the hyperbolic plane.
Figure \ref{fig:regular-graphs} shows two such graphs in panels (a) and (b) labeled with Sch{\"a}fli symbols $\{3,7\}$ and $\{3,8\}$, respectively, where the first element indicates the number of sides of each face and the second element indicates degree of each vertex. Panels (c) and (d) show the corresponding entropy curves as a function of subsystem size along with a fit to the single interval entropy formula. One observes that the single interval fit captures the overall shape of the entropy, but there is region-size-dependent substructure which is not captured. Looking at panels (a) and (b) and recalling the calculations in Section~\ref{sec:review}, one natural hypothesis is that the cutoff on the hyperbolic disk (the outer set of sites) is not at a uniform radius. In the language of the RT calculations above, the bulk cutoff $r_c$ is promoted to a function of $\theta$. In the context of the computation leading up to Equation~\ref{eq:ads_ent_cutoff}, if we imagine that the RT surface hits the cutoff surface at the midpoint between the region and its complement, this gives a simple prediction for $r_c(\theta)$ in terms of the scaled hyperbolic distance $d_{\textnormal{bulk}}$. Including this effect accounts for much of the substructure as seen in panels (c) and (d) of Figure \ref{fig:regular-graphs}. One could attempt to further refine this simple cutoff picture or allow for a general end-point-dependent offset in the entropy treated as a set of fitting parameters, but we do not pursue these directions here. For other figures in this supplement, we will often show results using the $\{3,7\}$ and $\{3,8\}$ graphs to complement the results obtained with the MERA graph.

% f the $i$th boundary point is located at $\theta = \theta_i$ and $r = r_c(\theta_i)$, then the entropy would of an interval from point $i$ to point $j$ would be modified to $S([i,j]) = S([i,j])_{\mathrm{uniform\,cutoff}} + s_i + s_j $ where $s_i \sim \ln \frac{r_c(\theta_i)}{r_{c,\mathrm{uniform}}}$.

% Additionally, a scaled hyperbolic distance $d_{\textnormal{bulk}}$ is shown that is based on the hyperbolic distance between midpoints of the boundary edges that are cut when producing the boundary subsystems. (These midpoints are a rough ansatz for the bulk positions of the boundary region endpoints, whose precise bulk positions are unknown because the exact dual picture is not specified.) The qualitative agreement between the scaled $d_{\textnormal{bulk}}$ and the entanglement entropies suggests that some of the small-scale variation in the entropies can be attributed to the variation in the bulk positions of the boundary region endpoints. The overall form for both graphs are reasonably fit by the RT formula for adequate choices of effective central charges particular to each graph.
% For other figures in this supplement, we will often show results using the $\{3,7\}$ and $\{3,8\}$ graphs to complement the results obtained with the MERA graph.

\section{Correlations and entanglement in the hyperbolic disk geometry}\label{sec:correlations-and-entanglement}

Our protocol for engineering holographic entanglement includes several tuning knobs: how strongly we squeeze the initial state, how finely we discretize the target bulk geometry in converting it to a graph, and whether or not we decorate the graph.  This section presents supplemental numerical data examining the dependence of correlations and entanglement on these parameters for the hyperbolic disk geometry.  We comment on implications for prospective experimental implementations and for the theoretical interpretation of our construction.

\subsection{Entanglement entropies}

\begin{figure}[!htbp]
    \centering
    \includegraphics[width=0.9\linewidth]{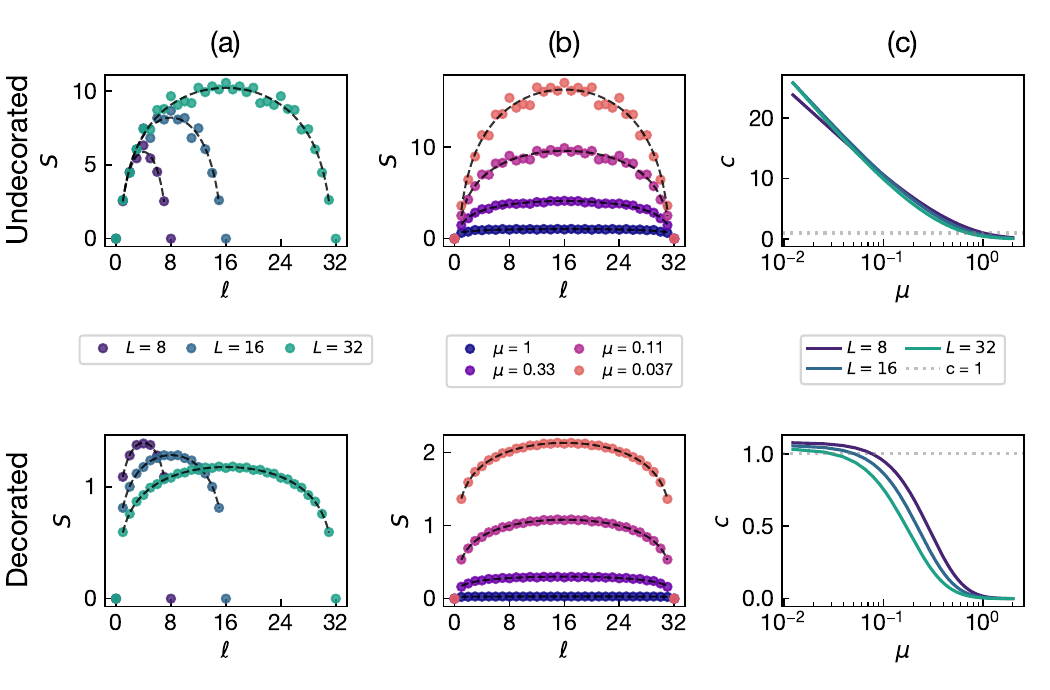}
    \caption{Comparison of results for different graph sizes and squeezing values for both the undecorated and decorated cases. The columns show (a) entanglement entropy for various boundary sizes (corresponding to
    depths of 3 through 5) with a fixed squeezing $\mu=0.1$, (b) entanglement entropy for various $\mu$ for fixed depth of 5, (c) fit central charge as a function of squeezing. For the entanglement entropy plots, fits to the (1+1)D CFT entanglement entropy curve are shown in dashed black lines. In the decorated case, in the limit of small $\mu$ the fit central charge $c$ approaches an approximate value of 1.}
    \label{fig:entanglement_entropy_comparison}
\end{figure}

While deep graphs are theoretically appealing for approaching the limit of a smooth geometry, smaller graphs offer the most convenient starting point for experimental implementations.   Thus, in Fig.~\ref{fig:entanglement_entropy_comparison}(a) we examine how the entanglement entropy $S$ depends on the size of the graph for depths $\depth_b = 3,4,5$ at fixed initial squeezing $\mu = 0.1$ in both the undecorated (top) and decorated (bottom) constructions.  Additionally, in Fig.~\ref{fig:entanglement_entropy_comparison}(b), we plot the entanglement entropy at fixed depth $\depth_b = 5$ for different values of the initial squeezing $\mu$.  In each case, the dependence of entanglement entropy $S$ on region size $\ell$ is well described by a fit with the functional form of the CFT prediction
\begin{equation}
    S_{\textnormal{gs}}(\ell) = \frac{c}{3} \ln \left( \sin \frac{\Delta \theta}{2}\right) + \epsilon,
\end{equation}
where $\Delta\theta = 2\pi \ell/L$.

From each fit to the entanglement entropy, we extract a central charge $c$.  In the undecorated case, we find that the central charge increases logarithmically with the inverse squeezing parameter, $c \propto \log(1/\mu)$, in the regime of strong squeezing $\mu \ll 1$.  By contrast, in the decorated case, the central charge approaches a saturation value $c \approx 1$ in the limit of strong squeezing $\mu \ll 1$.  Intriguingly, the saturation value $c\approx 1$ coincides with the prediction for a $1\!+\!1$-dimensional free boson CFT.

% \textcolor{red}{Somewhere, we need to discuss the two-region entanglement and link it to the figure~\ref{fig:rt-crossover-sharpness}!!}

While the limit of strong initial squeezing is a theoretically simple case in which quenching on the bulk couplings produces an ideal graph state, in practice experiments operate at finite squeezing and may contend with tradeoffs between the degree of squeezing and the complexity of control over the coupling graph.  Thus, the good agreement of the entanglement entropy with the CFT model regardless of initial squeezing --- and down to small system sizes --- offers convenience for near-term experimental implementations.

\begin{figure}[!htbp]
    \centering
    \includegraphics[width=0.75\linewidth]{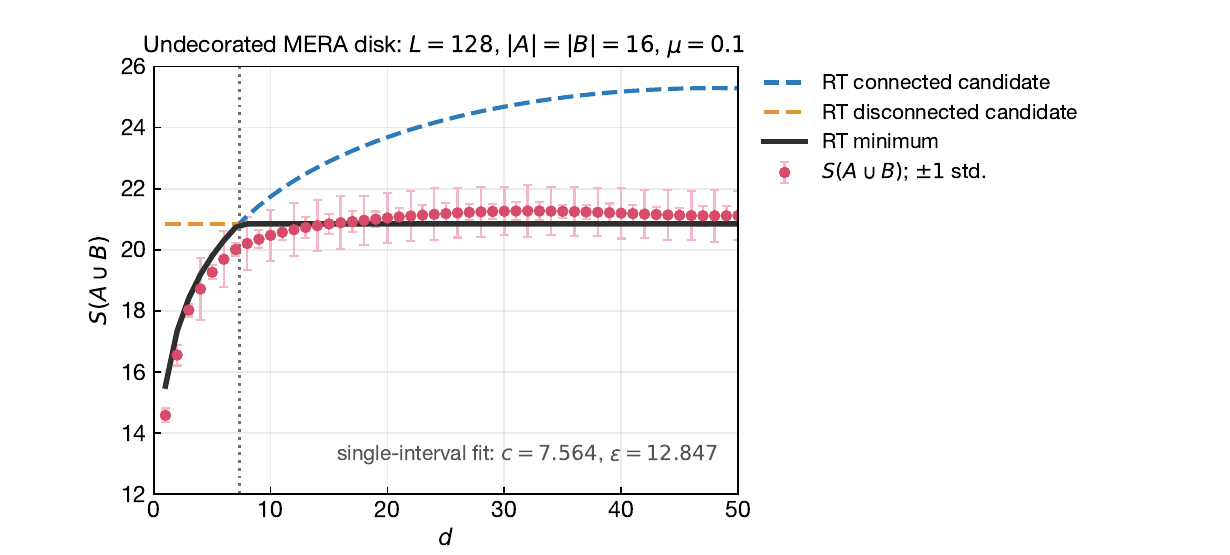}
    \caption{\diff{
    Two-region entanglement entropies for varying separation distance $d$ in the undecorated construction. There are two candidate RT surfaces to consider: one where the RT surfaces connect the two boundary regions, and another where the RT surfaces do not connect the two boundary regions. The calculated RT crossover does not sharply follow the transition between the two RT candidate surfaces, but shows fair agreement within the two regimes. The results are averaged over all initial points along the MERA boundary, and bars show $\pm1$ standard deviation of the entropy values.
    }}
    \label{fig:rt-crossover-sharpness}
\end{figure}

\diff{We can also study multi-region entanglement entropy. For example, Figure~\ref{fig:rt-crossover-sharpness} shows the two-region entanglement for the undecorated MERA graph as a function of the separation between the two intervals, $A$ and $B$, each of which is $16$ sites. The two dashed curves are the RT predictions obtained from two competing minimal surface configurations. We see that the actual data (red dots) are close the RT predictions for small and large separation, but the sharp phase transition predicted by RT is rounded (vertical dotted line). In the future, it will be interesting to better understand how well the undecorated state can reproduce such a transition, say at large size and small $\mu$. We also present some additional results for multi-region entanglement below when we discuss mutual information and tripartite information.}

\subsection{Limit of small $\mu$}

\begin{figure}[!htbp]
    \centering
    \includegraphics[width=0.75\linewidth]{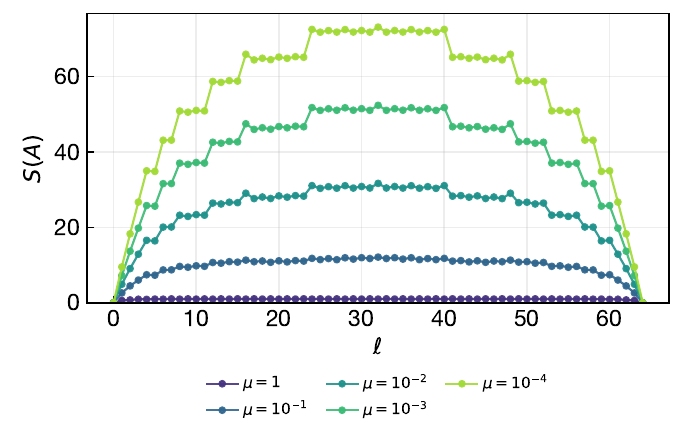}
    \caption{
    \diff{Entanglement entropy of a fixed undecorated periodic MERA graph as the squeezing parameter $\mu$ is decreased. For large $1/\mu$, the entropy curve develops pronounced lattice-scale structure and approaches a piecewise profile determined by the discrete coupling graph, rather than a smooth CFT-like function of boundary region size $\ell$.}
    }
    \label{fig:small-mu}
\end{figure}

Especially in the undecorated case, the entropy versus subsystem size curve develops more lattice-scale structure as $\mu$ is reduced, as in upper panel (b) of Figure~\ref{fig:entanglement_entropy_comparison}. Here we show that the undecorated entropy curve enters a qualitatively different regime when $1/\mu$ is made large. For a fixed graph at arbitrarily strong squeezing, the entropy curves become piecewise functions of the subregion that tend to jump as the subregion size is increased. Thus the entropy is no longer consistent with a smooth geometry and is determined instead by the detailed properties of the chosen coupling graph, as shown in Fig.~\ref{fig:small-mu}. 

To understand this, it is helpful to start with the coupling matrix which is determined by the graph adjacency matrix. If we separate the sites of the original pre-measured coupling graph into two sets, unmeasured $U$ and to-be-measured $M$, then the adjacency matrix can be written
\begin{equation}
    A = \begin{bmatrix}
        A_{UU} & A_{UM} \\ A_{MU} & A_{MM}
    \end{bmatrix}.
\end{equation}
The effect of the measurement is to generate a new effective weighted adjacency matrix for $U$ given by
\begin{equation}
    A_{\text{eff}} = A_{UU} - A_{UM} A_{MM}^{-1} A_{MU}
\end{equation}
where $A_{MM}^{-1}$ is the Moore-Penrose inverse in general. 

Now, when the squeezing is very large, we start by dropping $\mu x^2$ relative to $x A_{\text{eff}} x$ in the exponent of the wavefunction. In this limit, an elementary calculation reveals that the subsystem density matrix is some number of copies of a Dirac delta function with infinite entropy. This is the Gaussian analog of copies of the maximally mixed state. For a partition of $U$ into $U_A \cup U_B$, the number of copies is determined by the rank of $(A_{\text{eff}})_{U_A A_B}$. Adding back in the regulating effect of the squeezing term, the entropy is 
\begin{equation}
    S(U_A) = \text{rank}[(A_{\text{eff}})_{U_A A_B}] \ln \frac{1}{\mu} + \dots
\end{equation}
at large but finite $1/\mu$. Since the rank is an integer, we obtain the claimed step-function-like dependence of the entropy on subsystem size.

\subsection{Correlations and mutual information}

\begin{figure}[!htbp]
    \centering
    \includegraphics[width=\linewidth]{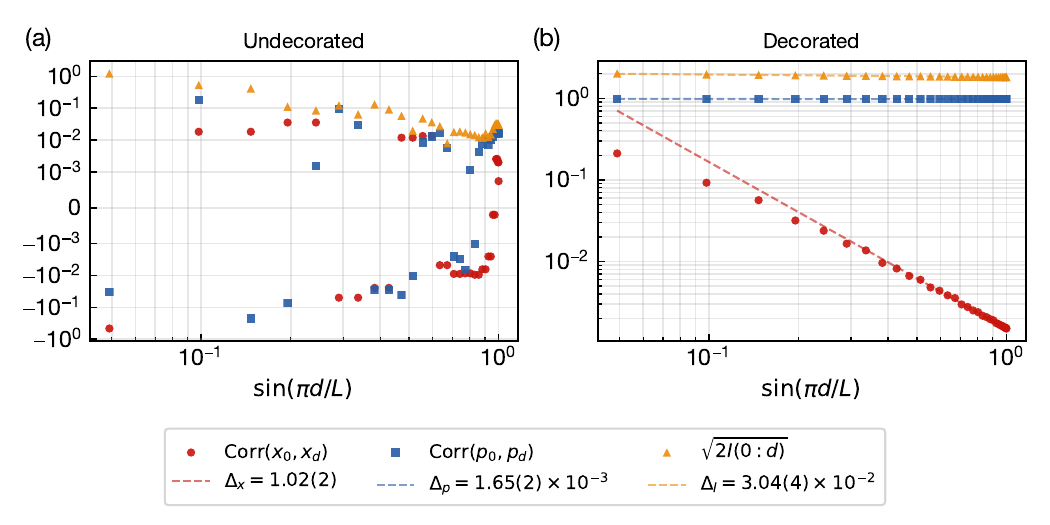}
    \caption{Correlations and mutual information in the (a) undecorated and (b) decorated case. For both cases, a graph with depth $R_b = 6$ is used with squeezing $\mu=0.01$. For the decorated case,
    power law fits (dashed lines) yield the respective scaling dimensions $\Delta$ shown in the legend. The least-squares fits are performed over the third of the points with the largest $\sin(\pi d / L)$, and the uncertainties shown are standard errors.}
    \label{fig:correlations}
\end{figure}

A major motivation for the decorated graph is that it produces correlations of uniform sign, allowing us to observe a clean power-law decay.  By contrast, analyzing the correlations in the undecorated case is complicated by their non-uniform sign.  However, a convenient proxy is to examine the mutual information $I(i:j)$ between pairs of sites $i,j$, which sets an upper bound on the correlations
\begin{equation}
\corr(\phi_i, \phi_j)=
\frac{\cov(\phi_i,\phi_j)}{\sqrt{(\var{\phi_i})(\var{\phi_j}})} \leq \sqrt{2 I(i:j)},
\end{equation}
where $\phi\in\{x,p\}$ is a quadrature operator.  The calculation of the mutual information and its role as a bound on covariances are reviewed in Sec.~\ref{sec:MI}.

\begin{figure}[!htbp]
    \centering
    \includegraphics[width=0.5\linewidth]{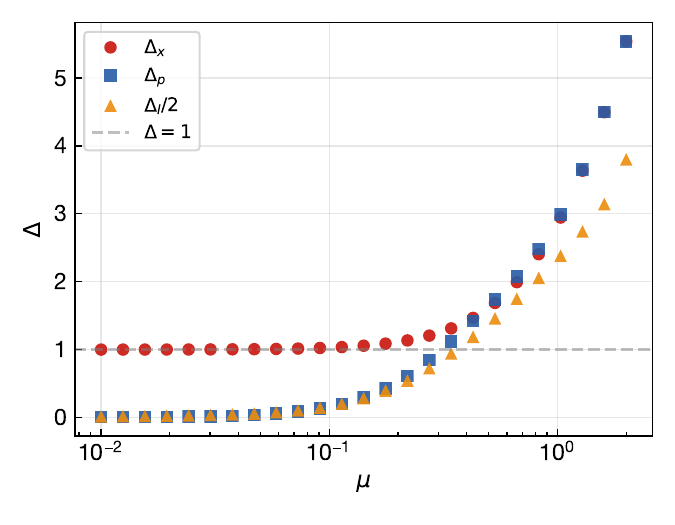}
    \caption{Scaling dimensions $\Delta$ for $x$, $p$, and mutual information $I$ as a function of squeezing $\mu$ for a decorated graph of depth $R_b = 6$.}
    \label{fig:decorated-scaling}
\end{figure}

We analyze correlations and mutual information obtained from the quench-and-measure protocol on the MERA graph in Fig.~\ref{fig:correlations}, which shows both the undecorated and decorated cases.
For both the undecorated and decorated cases, $\sqrt{2I(i:j)}$ indeed bounds both $x$ and $p$ correlations. Additionally, in the decorated case, power law behavior emerges at large $\sin(\pi d / L)$, where $d=|i - j|$ is the distance on the boundary separating the two nodes. We fit the correlations as a power law $\corr(\phi_0, \phi_d) \propto d^{-2\Delta_\phi}$ with the scaling dimension $\Delta_\phi$ for each operator $\phi$ as a free parameter. The corresponding exponent $\Delta_I$ for mutual information is fit by $I(\phi_0 : \phi_d) \propto d^{-2\Delta_I} $. These scaling dimensions depend on the squeezing $\mu$, as shown in Fig.~\ref{fig:decorated-scaling}. In particular, for small $\mu$, the scaling dimension $\Delta_x$ approaches 1. The domains for the fits are over the third of the data
with the largest $\sin(\pi d / L)$.

\subsection{\diff{Mutual information of two intervals}}

\begin{figure}[!htbp]
    \centering
    \includegraphics[width=0.68\linewidth]{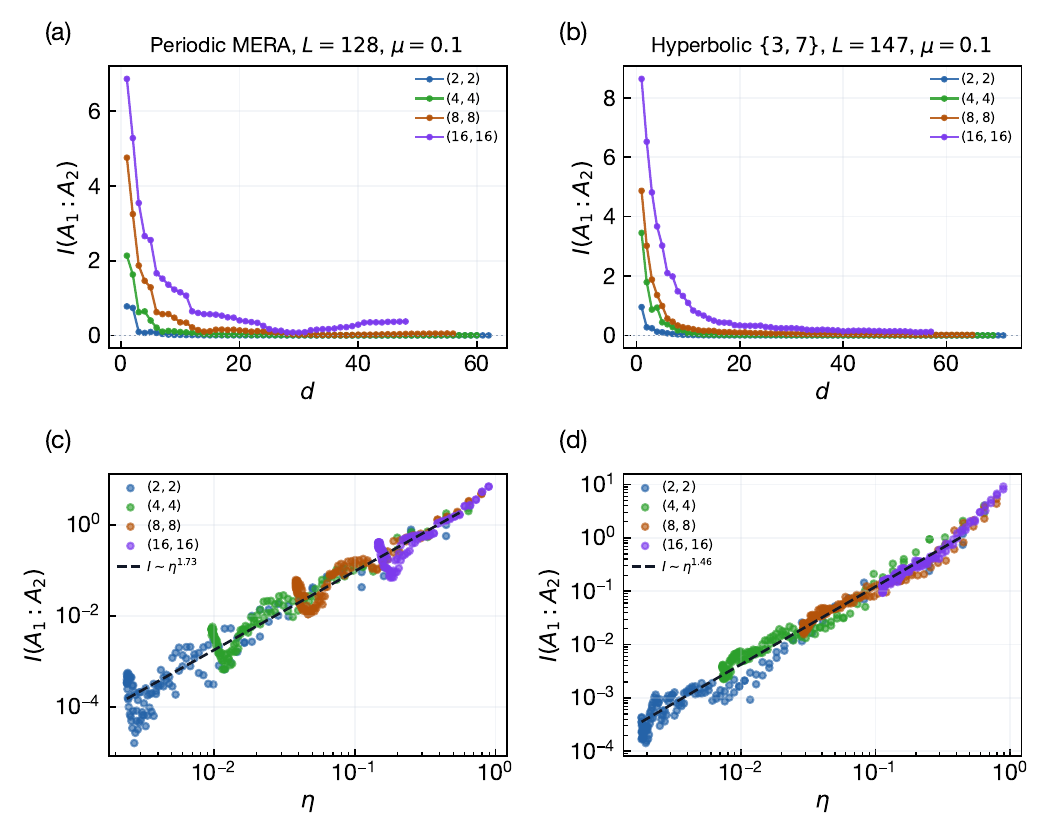}
    \caption{\diff{Boundary mutual information in the undecorated case for two regions of lengths $(\ell_1, \ell_2)$, plotting against separation distance $d$ for MERA (a) and a regular $\{3,7\}$ hyperbolic lattice (b). For (c) and (d), the MERA and $\{3, 7\}$ lattice mutual information are plotted against 
    the cross-ratio $\eta$.}
    }
    \label{fig:mera-undecorated-mutual-information}
    \vspace{0.5in}
    \includegraphics[width=0.68\linewidth]{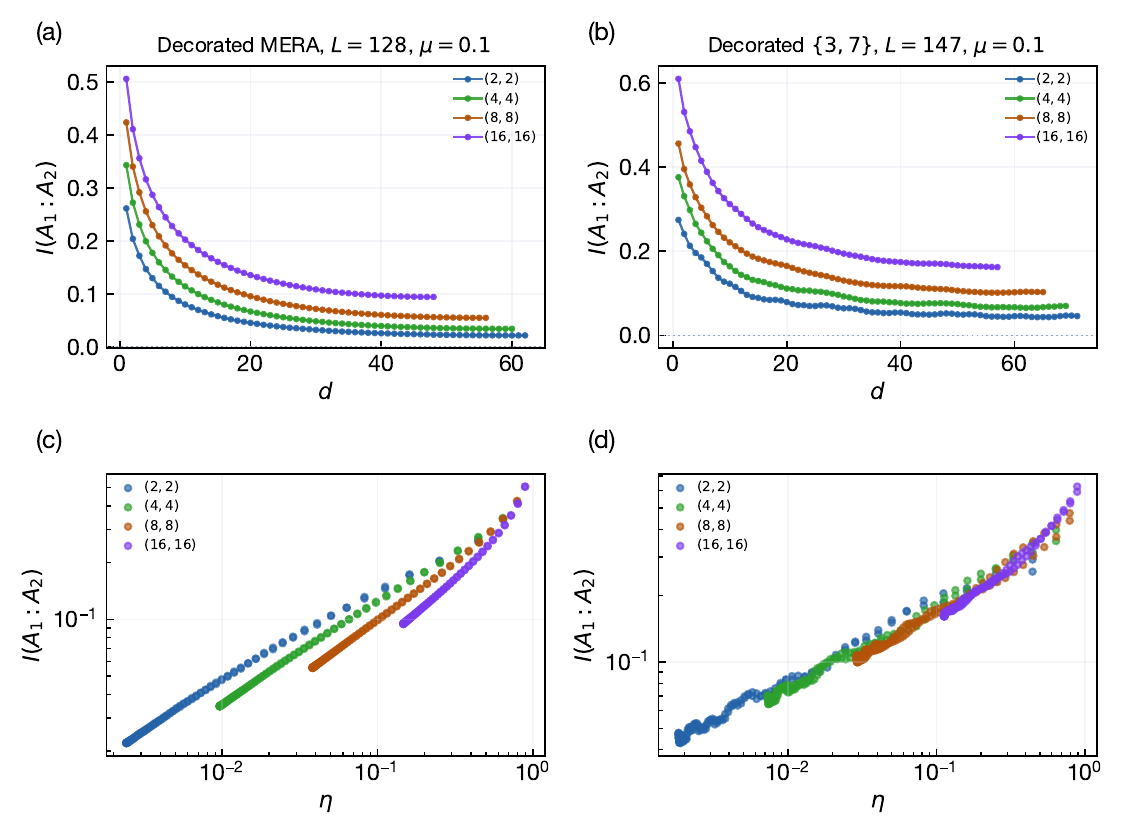}
    \caption{\diff{Boundary mutual information in the decorated case for two regions of lengths $(\ell_1, \ell_2)$, plotting against separation distance $d$ for MERA (a) and a regular $\{3,7\}$ hyperbolic lattice (b). For (c) and (d), the MERA and $\{3, 7\}$ lattice mutual information are plotted against 
    the cross-ratio $\eta$.}
    }
    \label{fig:mera-decorated-mutual-information}
\end{figure}

For a $(1+1)$d CFT, the mutual information $I(A_1:A_2)$ of two regions $A_1$ and $A_2$ exposes
information about the system beyond the central charge $c$. This motivates
examining the mutual information of two intervals of the boundary states
produced by our protocol. For non-overlapping regions of a $(1+1)$d CFT on a circle,
the mutual information simplifies to a function of the cross ratio
\begin{equation}
\eta = \frac{d(u_1,v_1)d(u_2,v_2)}{d(u_1, u2)d(v_1, v_2)}
\end{equation}
where $u_1$ and $u_2$ are endpoints of region $A_1$, $v_1$ and $v_2$ are
endpoints of region $A_2$, and $d(i,j)=\frac{L}{\pi}|\sin{\left( \pi (i - j) / L \right)}|$ gives the chord distance between points $i$ and $j$ on a
circle of circumference $L$.
In Fig.~\ref{fig:mera-undecorated-mutual-information}, we plot both $I(A_1:A_2)$ for the undecorated case both as a function of the boundary distance $d$ separating the two regions and the cross ratio $\eta$.
In Fig.~\ref{fig:mera-decorated-mutual-information}, we perform the same calculation for the decorated case.

\subsection{\diff{Tripartite information of three intervals}}

\begin{figure}[!htbp]
    \centering
    \includegraphics[width=\linewidth]{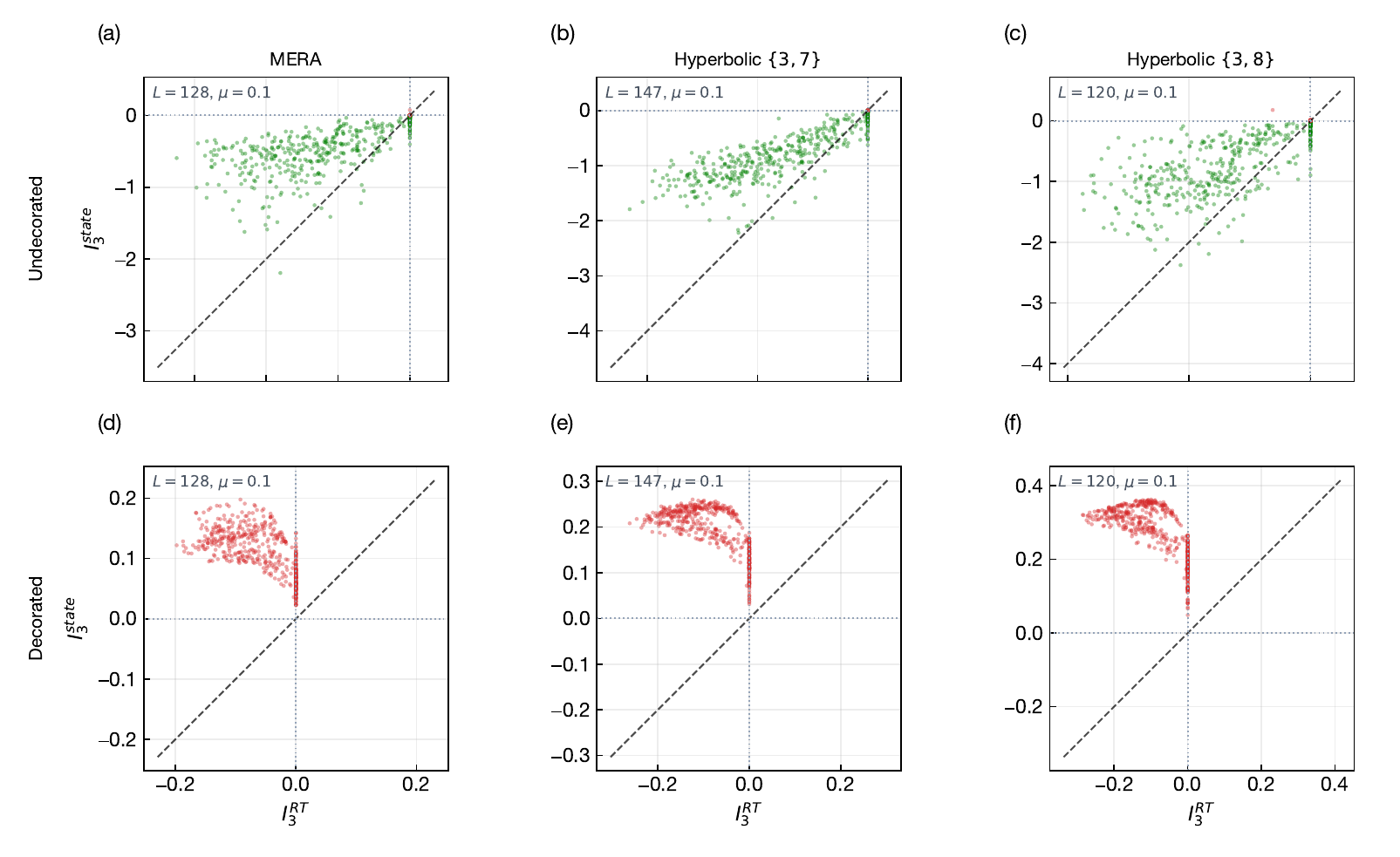}
    \caption{\diff{Tripartite information $I_3(A:B:C)$ for both the undecorated and decorated cases, sampled for 500 different interval triplets $A,B,C$ satisfying $|A|+|B|+|C| \geq 10$.  Here, we plot $I_3(A:B:C)$ for the state against the RT-prediction given the
    interval choices. Note that tripartite information is non-positive for states precisely obeying the RT formula. The non-positivity of $I_3$ is typically satisfied for the undecorated case, but
    is violated for the decorated case.
    }
    }
    \label{fig:tripartite-information}
\end{figure}

As an additional check, we can look at the tripartite information of three intervals, defined as
\begin{equation}
I_3 (A:B:C) = I(A:B) + I(A:C) - I(A:BC)
\end{equation}
For states that precisely obey the RT formula, $I_3(A:B:C) \leq 0 $ for all non-overlapping intervals $A$, $B$, $C$; this ``monogamy of mutual information'' (MMI) inequality \cite{hayden_holographic_2013} can be interpreted as regions $B$ and $C$ together providing more information about $A$ than $B$ and $C$ known separately. In Fig.~\ref{fig:tripartite-information}, we show the tripartite information for the
undecorated and decorated cases for both the MERA graph and regular graphs. We see that MMI is generally satisfied for the undecorated case, while it is violated for the decorated case. This indicates
that while the decorated construction produces a good approximation to a state of a CFT (as further justified by vector fixed point equation diagnostics in Sec.~\ref{sec:vfpe}), that CFT is not a holographic CFT. However, MMI being true in the undecorated case opens the door for a series of other holographic inequalities to possibly be satisfied. These higher-order holographic inequalities which involve more regions~\cite{Bao_2015}
can be checked, and we leave a more careful study to future work.

\section{Correlations, entanglement, and mutual information of Gaussian states}\label{sec:gaussian-states}

\subsection{Calculating entanglement entropies from the covariance matrix}
The entanglement structure of a Gaussian state is fully determined by the covariance matrix.  In the main text, after deriving the covariance matrix of the state prepared by the quench-and-measure protocol, we use it to determine the entanglement entropy $\SEE = -\Tr(\rho_A \log\rho_A)$ of a boundary subregion $A$ with density matrix $\rho_A$.  More generally, we evaluate R{\'e}nyi entropies
\begin{equation}\label{eq:Renyi}
S_\alpha = \frac{1}{1-\alpha}\log \Tr(\rho_A^\alpha),
\end{equation}
where the entanglement entropy is obtained in the limit $\SEE = \lim_{\alpha\rightarrow 1} S_\alpha$.  The calculation of entanglement entropy from covariances of Gaussian states was derived in Ref.~\mcite{holevo1999capacity} and is presented in several review articles~\mcite{eisert2010colloquium,weedbrook2012gaussian,demarie2018pedagogical}.  We nevertheless here provide a brief self-contained derivation aimed at providing physical intuition.

A foundation for our analysis is that any Gaussian state can be related by a \textit{symplectic transformation} to a product state whose entropy will be straightforward to calculate.  A symplectic transformation is one that preserves the commutation relations, and correspondingly also preserves the phase space volume and entropy.  By Williamson’s theorem~\mcite{williamson1936algebraic}, there exists such a transformation $\Sym$ that converts the covariance matrix $\Cmat_A$ to diagonal form: $\Sym \Cmat_A \Sym^T = D = \mathrm{diag}(\Seig_1, \Seig_1, \Seig_2, \Seig_2, \dots, \Seig_M, \Seig_M)$ for any $M$-mode Gaussian state.  In other words, $\Sym$ transforms to a basis of collective modes, in terms of which the system $A$ is in an unsqueezed product state $\rho_A = \otimes_m \rho_m$.  The \textit{symplectic eigenvalues} $\Seig_m$ represent the variances $\avg{X_m^2} = \avg{P_m^2} = \nu_m$ of quadrature operators in each collective mode indexed $m$, which is thus precisely in a thermal state with mean occupation $\avg{n}_m = \frac{1}{2}\avg{X_m^2 + P_m^2} - \frac{1}{2} = \Seig_m - \frac{1}{2}$, where the subtraction of 1/2 accounts for zero-point fluctuations.

We now use the statistical mechanics of the quantum harmonic oscillator to determine the entropy $S = -\avg{\log\rho}$ of each collective mode.  In particular, for a single mode in a thermal state of inverse temperature $\beta$, for which the mean occupation number is $\avg{n} = 1/(e^\beta - 1)$, the entropy is given by
\begin{equation}
S = \left(\avg{n}+1\right)\log\left(\avg{n}+1\right) - \avg{n} \log \avg{n}.
\end{equation}
Expressing the occupation $\avg{n_m} = \Seig_m-\frac{1}{2}$ of each eigenmode in terms of its symplectic eigenvalue $\Seig_m$ and summing over modes, we arrive at the entanglement entropy
\begin{align}\label{eq:SEE}
\SEE = \sum_m\left[\left(\Seig_m+\frac{1}{2}\right)\log\left(\Seig_m+\frac{1}{2}\right) - \left(\Seig_m-\frac{1}{2}\right) \log\left(\Seig_m-\frac{1}{2}\right)\right].
\end{align}
Note that Eq.~\ref{eq:SEE} correctly yields an entropy proportional to the logarithm of the phase space volume in the classical limit, since $\SEE \approx \sum_m (\log \nu_m + 1)$ when $\nu_m \gg 1$ for all modes.

The R{\'e}nyi entropies defined in Eq.~\ref{eq:Renyi} are analogously determined by first evaluating $\Tr(\rho^\alpha)$ for a thermal state $\rho$ of a harmonic oscillator.  In terms of the mean occupation $\avg{n}$, we obtain
\begin{equation}
\Tr(\rho^\alpha) = \frac{1}{(\avg{n}+1)^\alpha-\avg{n}^\alpha}.
\end{equation}
Substituting this result into Eq.~\ref{eq:Renyi} and summing over modes with symplectic eigenvalues $\Seig_m = \avg{n}_m+\frac{1}{2}$ yields R{\'e}nyi entropies
\begin{equation}\label{eq:Salpha}
S_\alpha = \frac{1}{\alpha-1}\sum_m \log\left[\left(\Seig_m+\frac{1}{2}\right)^\alpha-\left(\Seig_m-\frac{1}{2}\right)^\alpha\right].
\end{equation}

To make use of the above entropy formulas, we require a prescription for determining the symplectic eigenvalues from the correlation matrix.  Recall that the symplectic transformation $\Sym$, by definition, preserves the commutation relations $\Sform_{ij} = i[\xi_i,\xi_j]$. Here, $\boldsymbol{\xi} = (x_1, p_1, \dots x_M, p_M)$ is the vector of local quadrature operators, such that $\Sform = \oplus_{j=1}^M \Omega_j$ is the direct sum of matrices
\begin{equation}
\Sform_j = \begin{pmatrix} 0 & -1 \\ 1 & 0
\end{pmatrix}
\end{equation}
specifying the commutation relations for each oscillator indexed $j$.  Formally, then, a symplectic transformation is one that satisfies $\Sym \Sform\Sym^T = \Sform$.  From this symplectic identity, it can be shown that the diagonal matrix $\tilde{D}\equiv\Sform D$ has the same eigenvalues as $\tilde{\Cmat}_A\equiv\Sform\Cmat_A$.  In particular, both of these matrices have $2M$ eigenvalues $\lambda_{m\pm} = \pm i\Seig_m$.  Thus, finding the positive eigenvalues of $i\Cmat_A = i\Sform\Cmat_A$ by standard matrix diagonalization yields the symplectic eigenvalues $\Seig_m$ of the covariance matrix $\Cmat_A$.

We apply this symplectic diagonalization to covariance matrices $\Cmat_A$ of various subregions $A$ to calculate the entanglement entropies (Eq.~\ref{eq:SEE}) and R{\'e}nyi entropies (Eq.~\ref{eq:Salpha}) plotted in the main text.  The entropies thus calculated also allow us to determine mutual information $I(A:B) = S_A + S_B - S_{AB}$ between pairs of subsystems $A$ and $B$.

\subsection{Mutual information as a bound on correlations}\label{sec:MI}

We also comment that the mutual information bounds correlations of the oscillators. For bounded operators $O_A$ and $O_B$ supported on regions $A$ and $B$, there is a standard bound~\mcite{WolfVerstraeteHastingsCirac2008_MIcorr} on their covariance
\begin{equation}
    \cov(O_A,O_B) = \langle O_A O_B \rangle - \langle O_A \rangle \langle O_B \rangle,
\end{equation}
of the form
\begin{equation}
    \cov(O_A,O_B)^2 \leq 2 I(A:B) \| O_A \|^2 \| O_B \|^2.
\end{equation}
Position and momentum operators are unbounded, so this does not directly apply to them. 

For a pair of commuting variables which have a joint Gaussian probability distribution, $\xi_A$ and $\xi_B$, there is a well known classical bound
\begin{equation}
    \frac{\cov(\xi_A,\xi_B)^2}{\var(\xi_A) \var(\xi_B)} \leq 1 - e^{-2 I}.
\end{equation}
This follows by direct computation of the terms involved. Since $\xi_A$ and $\xi_B$ automatically commute if $A$ and $B$ are disjoint regions in the system, this classical bound suffices for bounding correlations between different regions of the boundary.

It is also true that $1-e^{-2I}\leq 2I$ for $I\geq 0$. They have identical values and first derivatives at $I=0$, and the second derivative of $1-e^{-2I}$ is less than zero. Thus $1-e^{-2I}$ grows more slowly than $2I$ and we also get the bound
\begin{equation}
    \frac{\cov(\xi_A,\xi_B)^2}{\var(\xi_A) \var(\xi_B)} \leq 2I.
\end{equation}
This is analogous to the general bound with the square of the operator norm replaced by the variance of the Gaussian variables.

\section{\protect\diff{Vector fixed point equation for the decorated graph}}\label{sec:vfpe}

In the main text we presented evidence that the boundary state produced by the decorated MERA graph was close to a conformal field theory (CFT) ground state. The scaling dimensions and central charges were approximately equal to those of a single free scalar field, a prototypical CFT. We also suggested that the state became a CFT state in the limit of large boundary size. We present evidence for that assertion here, using a tool known as the vector fixed point equation (VFPE)~\cite{lin2023conformal}.

Introduced in \cite{lin2023conformal}, given a quantum state, the VFPE demands the vanishing of a certain combination $K_\Delta$ of modular Hamiltonians $K_i$ obtained from the state. We define
\begin{equation}
K_{\Delta} \equiv \eta \hat{\Delta}(A, B, C)+(1-\eta) \hat{I}(A: C \mid B)
\end{equation}
where $A$, $B$, and $C$ are adjacent intervals, $\eta$ is the cross ratio of those intervals, and
\begin{equation}
\begin{aligned}
\hat{\Delta}(A, B, C) &\equiv K_{A B}+K_{B C}-K_A-K_C, \\
\hat{I}(A: C \mid B) &\equiv K_{A B}+K_{B C}-K_B-K_{A B C}.
\end{aligned}
\end{equation}
The VFPE condition is that
\begin{equation}
K_\Delta |\psi\rangle \propto |\psi\rangle.
\end{equation}
States that obey the VFPE equation are conjectured to be CFT ground states in $(1+1)$d. In practice, in finite lattice systems, a state will only approximately satisfy the VFPE, even if it is approaching a CFT state as the lattice size grows. It is useful to introduce the VFPE residual, chosen to be the standard deviation $\sigma(K_\Delta)$, to serve as the measure of the failure of the VFPE to be satisfied.

\begin{figure}[!htbp]
    \centering
    \includegraphics[width=\linewidth]{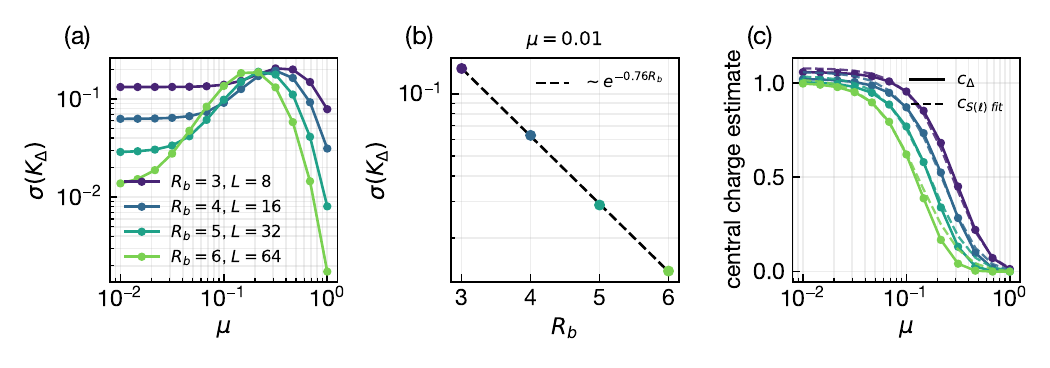}
    \caption{\diff{VFPE metrics for decorated MERA.
    In (a) the VFPE residual, quantified by $\sigma(K_\Delta)$, is shown as a function of $\mu$ for various lattice depths.
    In (b), $\sigma(K_\Delta)$ at a fixed $\mu$ of 0.01 is plotted against depth $R_b$ and fit against exponential decay.
     In (c) the $K_\Delta$-based estimate of the central charge is
     compared alongside the central charge fit using Eq.~\MainEqref{eq:cft-on-circle}, showing reasonable agreement across $\mu$. 
     }
    }
    \label{fig:vfpe}
\end{figure}

In Figure~\ref{fig:vfpe} we calculate the VFPE residual for decorated MERA graphs and find an exponential decay of the VFPE residual as a function of lattice depth. This provides additional evidence that the decorated graph produces a true CFT ground state in the limit of infinite size.

\StopSupplementTOCEntries

\end{document}